\documentclass[11pt, letterpaper, onecolumn, oneside, final]{article}

\usepackage{url}
\usepackage{times}
\usepackage{xcolor}
\usepackage{xspace}
\usepackage{amsthm}
\usepackage{amssymb}
\usepackage{amsmath}
\usepackage{enumitem}
\usepackage{graphicx}
\usepackage{makecell}
\usepackage{booktabs}
\usepackage{etoolbox}
\usepackage{algorithm}
\usepackage[compact]{titlesec} % less spacing on titles
\usepackage[noend]{algpseudocode}
\usepackage{thmtools, thm-restate}
\usepackage[sort,compress,numbers]{natbib}
\usepackage[font=footnotesize,labelfont=bf]{caption} % figure captain is smaller and bold
\usepackage[left=1in,right=1in,top=1in,bottom=1in]{geometry} % set up page margins
\usepackage[bookmarks=true,bookmarksnumbered=true]{hyperref}
\usepackage{cleveref}

\theoremstyle{plain}
\newtheorem{theorem}{Theorem}[section]
\newtheorem{definition}[theorem]{Definition}
\newtheorem{fact}[theorem]{Fact}

\newtheorem{claim-inline}{Claim}[theorem]

\newtheorem{property}[theorem]{Property}
\newtheorem*{claim*}{Claim}

\newtoggle{submission}

\toggletrue{submission}
\iftoggle{submission}
{
	
	\newcommand{\highlightblue}[1]{{\color{black} #1}}
	\newcommand{\highlight}[2][black]{{\color{black} #2}}
}{
	
	\newcommand{\highlightblue}[1]{{\color{blue} #1}}
	\newcommand{\highlight}[2][red]{{\color{#1} #2}}
}

\iftoggle{submission}
{

}{

}

\makeatletter
\algnewcommand{\LineComment}[1]{\Statex\hskip\ALG@thistlm $\blacktriangleright$ #1}
\makeatother

\newcommand{\id}{\texttt{ID}}
\newcommand{\nid}{\texttt{NewID}}
\newcommand{\hash}{\texttt{Hash}}
\newcommand{\validator}{\texttt{Validator}}
\newcommand{\consensus}{\texttt{Consensus}}

\begin{document}

%%% cover page

\title{\textbf{Robust and Scalable Renaming with Subquadratic Bits}}
\author{
Sirui Bai\\
Nanjing University
\and
Xinyu Fu\\
Nanjing University
\and
Yuheng Wang\\
TU Wien
\and
Yuyi Wang\\
CRRC Zhuzhou Institute \& Tengen Intelligence Institute
\and
Chaodong Zheng\\
Nanjing University
}
\date{
%\today
}
\maketitle
\thispagestyle{empty}

\begin{abstract}
In the renaming problem, a set of $n$ nodes, each with a unique identity from a large namespace $[N]$, needs to obtain new unique identities in a smaller namespace $[M]$. A renaming algorithm is strong if $M=n$, and is order-preserving if the relative order between identities is preserved after execution.
Renaming is a classical problem in distributed computing with a range of applications, and there exist many time-efficient solutions for fault-tolerant renaming in synchronous message-passing systems. However, all previous algorithms send at least $\Omega(n^2)$ messages, and many of them also send large messages each containing $\Omega(n)$ bits. Moreover, most algorithms' performance metrics do not scale with the actual number of failures. These limitations restrict their practical performance.

In this paper, we develop two new strong renaming algorithms, one tolerates up to $n-1$ crash failures, and the other tolerates up to $(1/3-\epsilon_0)n$ Byzantine failures for an arbitrarily small constant $\epsilon_0>0$.
The crash-resilient algorithm is always correct and always finishes within $O(\log{n})$ rounds. It sends $\tilde{O}((f+1)\cdot n)$ messages with high probability in $n$, where $f$ is the actual number of crashes during execution. This implies that it sends subquadratic messages as long as $f=o(n/\log{n})$.
The Byzantine-resilient algorithm trades time for communication cost. Specifically, it finishes within $\tilde{O}(\max\{f,1\})$ rounds and sends only $\tilde{O}(f+n)$ messages, with high probability in $n$. Here, $f$ is the actual number of Byzantine nodes during execution. This algorithm is also order-preserving. To obtain such strong guarantees, the Byzantine-resilient algorithm leverages shared randomness and message authentication.
Both algorithms only send messages of size $O(\log{N})$ bits. Therefore, our crash-resilient algorithm incurs $o(n^2)$ communication cost as long as $f=o(n/(\log{n}\log{N}))$; and our Byzantine resilient algorithm incurs almost-linear communication cost.
By deriving a lower bound showing that any randomized strong renaming algorithm that succeeds with probability at least $3/4$ sends $\Omega(n)$ messages in expectation, we conclude that our algorithms achieve near-optimal communication cost in many cases.
\end{abstract}

\iftoggle{submission}
{}
{
\vspace{3em}\begin{Huge}
\begin{center}
{\bf\highlight{
!!! DO NOT EDIT THIS FILE !!! \\
!!! This file is for final merge !!!
}}
\end{center}
\end{Huge}
}

\clearpage
\thispagestyle{empty}
\tableofcontents

\clearpage
\pagestyle{plain}
\setcounter{page}{1}

% % % main content starts here % % %

\section{Introduction}\label{sec:intro}

\emph{Renaming} is a fundamental problem in distributed computing that involves symmetry breaking~\cite{attiya90}. At its core, renaming requires participating nodes with original identities to select new identities from a smaller range.
%Specifically, the renaming problem requires $n$ different nodes, each with an original identity from a large namespace $[N]=\{1,\cdots, N\}$, to select new identities from a target namespace $[M]$, where $n \leq M \ll N$.
This problem is relevant since the size of the nodes' namespace can affect the performance of many distributed algorithms, such as vertex coloring algorithms~\cite{barenboim13}, network decomposition algorithms~\cite{ghaffari2021improved}, and the recent MST algorithm focusing on energy complexity~\cite{augustine2024awake}.
Renaming is also helpful for practical distributed systems such as cryptocurrency networks, as using nodes' original identities for communication can be costly due to the varying network domains where the participants come from~\cite{bonneau15}.

To address the renaming problem, the most crucial requirement is \emph{uniqueness}, which demands that all participating nodes obtain distinct new identities. We now give a formal definition of this problem:

\begin{definition}[Renaming]\label{def:renaming}
Consider a distributed system consisting of $n$ nodes. Each node $v$ is assigned with an original unique identity $\id(v)\in[N]=\{1,\cdots,N\}$. That is, for any two nodes $u\neq v$, it holds that $\id(u)\neq\id(v)$. In the \emph{renaming} problem, each node $v$ needs to select a new unique identity $\nid(v)\in[M]$, where $n\leq M< N$. That is, for any two nodes $u\neq v$, it holds that $\nid(u)\neq\nid(v)$.
\end{definition}

Besides uniqueness, renaming algorithms may provide additional desirable properties. One such property is \emph{strong renaming} (also called \emph{tight renaming}), which ensures that the size of the target namespace exactly matches the number of participating nodes (that is, $M=n$). Another favorable property is \emph{order-preserving}, which preserves the relative order of identities after renaming (that is, $\id(u)<\id(v)$ iff $\nid(u)<\nid(v)$). Order-preserving is helpful when identities encode additional information, such as nodes' priorities when accessing a shared resource.

Renaming has been extensively studied by the community. (See, e.g., \cite{castaneda11,alistarh15survey} for a survey.) In this paper, we focus on fault-tolerant and efficient renaming in synchronous message-passing systems.

Specifically, we consider a fully connected network consisting of $n$ nodes; each node $v$ knows its own unique identifier $\id(v)$ and the value of $n$. All nodes are activated simultaneously and exchange messages in synchronous rounds. Each message accommodates at most $\Theta(\log{N})$ bits, where $N$ is the size of the original namespace.
Some nodes may fail during execution, but the number and the identities of the failed nodes are unknown. We consider two types of failures, namely \emph{crash failure} and \emph{Byzantine failure}.

A node that suffers crash failure effectively aborts execution and will not send any messages. We allow a node to crash at an arbitrary time, even in the middle of sending a message. We assume an \emph{adaptive adversary} called Eve determines when and which nodes crash. The adaptivity of Eve is characterized by the ability to use execution history up to any specific time point to decide which nodes crash immediately.

Byzantine failure is a much stronger type of failure: a Byzantine node can arbitrarily deviate from the specified protocol. For example, a Byzantine node can send arbitrary messages, or simulate crash failure. We assume a \emph{static adversary} called Carlo determines which nodes are Byzantine before the nodes are activated. To highlight our strategy for dealing with Byzantine failure, we also assume nodes can access shared random bits and messages are authenticated (so that nodes cannot spoof messages or identities). While these assumptions may seem strong, they are feasible in practice, and some could be weaken or even removed at the cost of a more complicated algorithm, see \Cref{subsec:byz-discussion} for more discussion.

As mentioned earlier, renaming is a well-studied problem. Hence, many solutions exist for settings similar or identical to ours, such as \cite{chaudhuri99,okun10,alistarh14} for the crash failure scenario, and \cite{okun2006renaming,okun08,denysyuk13} for the Byzantine failure scenario. The primary focus of these works has been time complexity. In particular, many proposed algorithms have poly-logarithmic running time, while the fastest one uses only sub-logarithmic rounds~\cite{alistarh14}. In spite of these exciting results, limitations still exist that constrain their practical performance:

\paragraph{High communication cost.} To the best of our knowledge, all existing renaming algorithms designed for the message-passing model involve some form of all-to-all communication. That is, during the execution of these algorithms, every node sends at least one message to every other node. Consequently, the message complexity of all such algorithms is $\Omega(n^2)$. Moreover, many algorithms (such as \highlightblue{\cite{okun2006renaming,okun08,okun10,alistarh12,denysyuk13b}}) require sending large messages each of which containing $\Omega(n)$ bits, resulting in $\Omega(n^3)$ bit complexity. Such high communication cost would hinder the scalability of these solutions, rendering them unsuitable for large networks. Although the inevitability of such complexity has been demonstrated for asynchronous crash-prone setting~\cite{alistarh15}, it remains unknown whether the quadratic bound can be broken in less harsh scenarios.

\paragraph{Inadaptive against severity of failure.} Most existing works on fault-tolerant renaming focus on attaining good worst-case performance. Hence, the resulting algorithms---while offering strong robustness---are unable to adapt their complexity to the severity of failures. To the best of our knowledge, the only exceptions are \cite{alistarh12,alistarh14}, which adjust the running time according to the actual number of failures. In practice, faults due to internal error may rarely occur, and external adversaries may not fully exploit their resources to corrupt all nodes under their control. Consequently, algorithms designed solely for the worst-case scenario could incur unnecessary resource expenditure and reduce practical efficiency.

In conclusion, these drawbacks raise two natural and important questions: (1) \emph{Can we design renaming algorithms that eliminate all-to-all communication?} (2) \emph{Can we design renaming algorithms whose complexity scale with the actual number of failures?}

\subsection{Results and Contributions}

We give affirmative answers to both questions by providing two randomized strong renaming algorithms: one achieving crash-resilience and one achieving Byzantine-resilience. We also devise a lower bound on the message/bit complexity of renaming, demonstrating our solutions are near-optimal in many settings.

\paragraph{Crash-resilient renaming algorithm.}

The guarantees enforced by our crash-resilient renaming algorithm is summarized in the following theorem. Note that its correctness and time complexity hold deterministically; while its message/bit complexity scale with the actual number of failures.

\begin{theorem}[Crash-resilient algorithm]\label{thm:main-crash}
In the synchronous message-passing model, there exists a strong renaming algorithm that is always correct and always terminates within $O(\log{n})$ rounds. It sends $O((f+\log n)\cdot{n}\log{n})$ messages with high probability in $n$, and never sends more that $\Theta(n^2\log{n})$ messages.\footnote{An event happens with high probability in $n$ if it occurs with probability at least $1-1/n$.} Here, $f<n$ is the actual number of crashes during execution and every message is of $O(\log{N})$ bits.
\end{theorem}

A direct implication of the above theorem is, so long as the actual number of failures is $o(n/\log{n})$, the total number of messages sent by our algorithm is $o(n^2)$. Moreover, since each message is of $O(\log{N})$ bits, as long as $N$ is polynomial in $n$, our algorithm works in the standard CONGEST model and has $O((f+\log{n})\cdot n\log^2{n})$ bit complexity, which is subquadratic when $f\in o(n/\log^2{n})$.

In designing the algorithm, to address the issue of high communication cost, we eliminate all-to-all communication by creating a ``committee''. Specifically, during algorithm execution, a limited number of nodes will be elected as committee members. {Rather than broadcasting messages to all nodes, each node communicates solely with the committee members and waits for their decisions. As long as the committee size is small, low message complexity can be achieved.

Nevertheless, during algorithm execution, both normal nodes and committee members can be crashed by the adversary. Such failures may result in conflicts in committee members' decisions or even the abortion of the entire algorithm in case all committee members crash. We solve these problems through conflict resolution mechanisms and committee re-election schemes, the latter of which is the key to achieving adaptivity against the actual number of failures. Specifically, our algorithm triggers the committee re-election procedure only when necessary, rather than in every round. Moreover, the timing and the probability for remaining nodes to join the committee are carefully designed to balance efficiency and robustness.

Within the committee, renaming is done through ``interval halving'' (see, e.g., \cite{okun08}). In this approach, each node maintains an interval that is initially set to $[1,n]$. At any point during execution, this interval specifies the range from which the node selects its new identity. The algorithm progresses through multiple phases; in each phase, nodes will try to halve their intervals with the help of committee members. Once the interval for a node is reduced to the size of one, the node has obtained a new identity in $[n]$.

\paragraph{Byzantine-resilient renaming algorithm.}

The guarantees enforced by our Byzantine-resilient renaming algorithm is summarized in the following theorem.
%Note that in exchange for stronger fault-tolerance, our algorithm relies on some additional assumptions when compared with the crash-tolerant scenario.
Note that similar to the crash scenario, as long as $N$ is polynomial in $n$, the Byzantine-resilient algorithm works in the standard CONGEST model. In fact, the bit complexity of this algorithm is almost linear (particularly, within poly-logarithmic factor), allowing it to scale nicely as the network size grows.
%\highlight{Moreover, to the best of our knowledge, this is the first algorithm that could achieve $o(n)$ time complexity with both strong and order-preserving properties in the synchronous message-passing model with Byzantine failures.}\marginnoteviolet{CD: previous strong and order-preserving Byzantine algorithms always use $\Omega(n)$ rounds?}

\begin{theorem}[Byzantine-resilient algorithm]\label{thm:main-byz}
In the synchronous message-passing model, assuming messages are authenticated and nodes can access shared random bits, there exists a strong and order-preserving renaming algorithm that succeeds within $O(\max\{f\log{N},1\}\cdot\log{n})$ rounds and sends \highlight{$O(f\cdot\log{N}\cdot\log^3{n} + n\cdot\log{n})$} messages, with high probability in $n$.
Here, $f<(1/3-\epsilon_0)n$ is the actual number of Byzantine nodes and $\epsilon_0>0$ is an arbitrarily small constant; each message is of $O(\log{N})$ bits.
\end{theorem}

Our Byzantine-resilient renaming algorithm also relies on using a committee to reduce communication cost. Nonetheless, with the help of shared random bits, the committee election process becomes simpler.\footnote{Electing and maintaining a small committee consisting of mostly correct nodes with limited cost is possible, but highly non-trivial, see \cite{augustine20} for an example. In this paper, we simplify this process with shared randomness as it is not our main objective.} The main challenge we have to overcome is to deal with the much stronger level of inconsistency caused by Byzantine failures, without using too much time and too many messages.

To begin with, observe that the interval halving approach used in the crash scenario is no longer feasible. This is because the adversary can manipulate the messages Byzantine nodes sent so that committee members would send incorrect decisions to nodes, resulting in duplicate new identities. \highlightblue{To avoid such issue, some form of consensus on the nodes' current intervals must be reached, and this is what previous work~\cite{okun08} has done.} Unfortunately, this would cost too much time and communication, even within a small committee.

Our algorithm takes a different approach: the committee tries to agree on a size $N$ bit vector denoting which nodes are currently present. Note that committee members cannot directly exchange these bit vectors, as that would again cost too much communication. Instead, committee members exchange hashes (i.e., ``fingerprints'') of their respective bit vectors to reduce complexity. However, this technique creates a new problem: if committee members discover inconsistency among hash values---which is inevitable due to failures---they cannot determine which locations in their bit vectors differ.

To address this new issue, our algorithm takes a divide-and-conquer approach on reaching consensus on bit vectors. More specifically, whenever the discrepancy among the hashes of the bit vectors become too large so that consensus is impossible, the bit vectors are divided into two segments of half size, and the process recurse. Since the number of Byzantine nodes is limited, the overall complexity of this process can be bounded and scales with the actual number of failures. We stress that though the high-level idea is not complex, implementing it correctly and efficiently is non-trivial. In particular, we employ both classical consensus and variants such as weak validator~\cite{lenzen22} to strike balance between correctness and efficiency.

%\highlightblue{We believe the approach of combining fingerprinting with divide-and-conquer could be of independent interest, and might be helpful for obtaining efficient Byzantine-resilient algorithms for other problems.}
We believe the approach of combining fingerprinting with divide-and-conquer might be helpful for obtaining communication efficient Byzantine-resilient algorithms for other problems.

\paragraph{Lower bound.}

To complement algorithmic results, we also derive an $\Omega(n)$ lower bound on the message/bit complexity for the strong renaming problem.
It shows our Byzantine-resilient algorithm achieves near-optimal (within poly-logarithmic factor) message/bit complexity. This lower bound is strong in that it holds even when nodes have shared randomness, messages are authenticated, and no failures occur.

\begin{restatable}{theorem}{ThmLowerBound}\label{thm:lower-bound}
In the synchronous message-passing model, any randomized strong renaming algorithm starting with an original namespace $N\geq 5n^2$ and succeeding with probability at least $3/4$ sends $\Omega(n)$ messages/bits in expectation, even with shared randomness and authenticated messages.
\end{restatable}

To prove above result, we first define the anonymous renaming problem. Then, we argue that if an anonymous renaming algorithm sends few messages, then two nodes have to select new identities without any communication, resulting in a non-trivial probability for them to pick identical new identities, hence obtaining a lower bound. Finally, apply a reduction and we are done.
See Appendix~\ref{secapp:lower-bound} for complete proof.

\subsection{Related work}\label{subsec:related-work}

Renaming was first introduced in \cite{attiya90}, where the authors showed the impossibility of tight renaming in asynchronous message-passing systems with crash failures and proposed an algorithm for a weaker version instead. Since then, renaming has been widely studied in both the message-passing model and the shared-memory model. In this part, we focus on discussing related work in the synchronous message-passing model. Interested readers can refer to surveys such as \cite{castaneda11,alistarh15survey} for more results.

Early results on renaming in the synchronous message-passing model often rely on consensus~\cite{lamport82} and reliable broadcast~\cite{bracha85}, and the round complexity of these algorithms often grow linearly with the maximum number of faults~\cite{dolev82}. Since it is believed that renaming is ``weaker'' than consensus~\cite{denysyuk13}, more recently, new algorithms have been proposed that leverage more specific tools and techniques.

For instance, in~\cite{okun08}, a Byzantine-resilient strong renaming algorithm based on interval halving is proposed by Okun, Barak, and Gafni.
This algorithm has $O(\log{n})$ time complexity, but sends $\Omega(n^2\log{n})$ messages each of $\Omega(n\log{N})$ bits. Our crash-resilient renaming algorithm also uses the interval-halving approach but has reduced message and bit complexity. In 2010, Okun proposed another crash-resilient renaming algorithm based on approximate agreement, which has $O(\log{n})$ time complexity~\cite{okun10}. Though algorithms in \cite{okun08,okun10} are both deterministic ones, randomization has also been adopted in solving renaming. For example, Denysyuk and Rodrigues proposed a randomized algorithm that employs cryptography tools to solve strong renaming~\cite{denysyuk13}. This algorithm can tolerate $n-1$ Byzantine failures. In \cite{alistarh14}, the authors constructed a tree structure and used the idea of balls-into-bins load-balancing to deal with renaming. Our Byzantine-resilient renaming algorithm uses shared randomness to construct hash functions, enabling nodes to roughly agree on a large bit vector while exchanging only limited number of bits.

\Cref{tbl:results} summaries most relevant existing results, along with our new results. To simplify presentation, we use $\tilde{O}(\cdot)$ to hide poly-logarithmic factors in $n$ and $N$.
%As can be seen, all previous work involve some form of all-to-all communication, and many of them rely on using large messages of $\Omega(n)$ bits. In contrast, our renaming algorithms eliminate the need for global broadcast and only use small messages. Moreover, our crash-resilient algorithm has accomplished such goal without sacrificing time complexity; while our Byzantine-resilient algorithm could attain poly-logarithmic time complexity if messages of size $\tilde{\Omega}(n)$ bits are permitted.

\begin{table}[t]
\centering
\begin{small}
\begin{tabular}{ccccccc}
\toprule
& Fault-tolerance & Runtime & \makecell{Total\\ Messages} & \makecell{Total\\ Bits} & Strong & \makecell{Order\\ Preserving} \\
\midrule
Chaudhuri et al.~\cite{chaudhuri99} & Crash ($f<n$) & $O(\log n)$ & $\tilde O(n^2)$ & $\tilde O(n^2)$ & yes & - \\
Okun~\cite{okun10} & Crash ($f<n$) & $O(\log n)$ & $\tilde O(n^2)$ &  $\tilde O(n^3)$ & yes & yes \\
Alistarh et al.~\cite{alistarh12} & Crash ($f<n$) & $O(\log f)$ & $\tilde O(n^2)$ & $\tilde O(n^3)$ & yes & yes \\
Alistarh et al.~\cite{alistarh14} & Crash ($f<n$) & $O(\log \log f)$ & $\tilde O(n^2)$ & $\tilde O(n^2)$ & yes & - \\
This work & Crash ($f<n$) & $O(\log n)$ & $\tilde O(fn)$ & $\tilde O(fn)$ & yes & - \\
\midrule
Okun et al.~\cite{okun2006renaming} & Byzantine ($f<n/3$) & $O(n)$ & $O(n^3)$ & $\tilde{O}(n^3)$ & yes & yes \\
Okun et al.~\cite{okun2006renaming} & Byzantine ($f<n/3$) & $O(\log n)$ & $\tilde O(n^2)$ & $\tilde O(n^3)$ & - & - \\
Okun et al.~\cite{okun08} & Byzantine ($f<n$) & $\tilde O(n)$ & $\tilde O(n^2)$ & $\tilde O(n^2)$ & yes & - \\
Okun et al.~\cite{okun08} & Byzantine ($f<n/3$) & $O(\log n)$ & $\tilde O(n^2)$ & $\tilde O(n^3)$ & - & - \\
Denysyuk et al.~\cite{denysyuk13} & Byzantine ($f<n$) & $O(\log n)$ & $\tilde O(n^2)$ & $\tilde O(n^2)$ & yes & - \\
Denysyuk et al.~\cite{denysyuk13b} & Byzantine ($f<n/3$) & $O(\log n)$ & $\tilde O(n^2)$ & $\tilde O(n^3)$ & - & yes \\
This work & Byzantine ($f<(1/3-\epsilon_0)n$) & $\tilde O(f)$ & $\tilde O(f+n)$ & $\tilde O(f+n)$ & yes & yes \\
\bottomrule
\end{tabular}
\end{small}
\vspace{-1ex}
\caption{Comparison of existing fault-tolerant renaming algorithms in the synchronous message-passing model.}\label{tbl:results}
\vspace{-2ex}
\end{table}

As can be seen, besides low complexity, our algorithms also adapt their costs to the actual severity of failures. This can be seen as an application of \emph{resource-competitive analysis}~\cite{gilbert12,bender15}. A resource-competitive algorithm should relate its performance metrics (such as time or message complexity) to the resources spent by the adversary on disrupting algorithm execution (such as the number of nodes the adversary corrupt). Hence, resource-competitive algorithms tend to perform better than algorithms that focusing on optimizing worst-case complexity. In recent years, resource-competitive analysis has been applied in many different scenarios, such as robust wireless communication~\cite{gilbert12making,gilbert14,king18,chen21}, noisy tolerance~\cite{dani18}, and Byzantine agreement~\cite{augustine20}. Prior to this work, only \cite{alistarh12,alistarh14} incorporate resource-competitiveness into renaming.

\section{Crash-resilient Renaming Algorithm}

\subsection{Algorithm Description}

Upon activation, each node $v$ initializes its interval $I_v=[1,n]$, and sets $p_v=0$, $d_v=0$, and $elected_v=false$. Here, $p_v$ controls the probability for $v$ to become a committee member (specifically, $v$ becomes a committee member with probability $\Theta((2^{p_v}\cdot\log{n})/n)$), $d_v$ is used to inform other nodes regarding the size of $I_v$ (roughly speaking, $|I_v|\approx n/2^{d_v}$), and $elected_v$ indicates whether $v$ is currently a committee member.
During initialization, each node elects itself as a committee member with probability $(256\log{n})/n$.

Note that although keeping both $d_v$ and $I_v$ seems redundant, we track these two distinct notations to simplify presentation. Specifically, imagine a binary tree in which each vertex is labeled with an interval; particularly, the root is labeled with interval $[1,n]$. Consider a vertex labeled with some interval $I=[l,r]$ in the tree. If $I$ contains more than one integer, then the vertex has two children: the left child is labeled with interval $\texttt{bot}(I)=[l,\lfloor(l+r)/2\rfloor]$, while the right child is labeled with interval $\texttt{top}(I)=[\lfloor(l+r)/2\rfloor+1,r]$. In our algorithm, $d_v$ equals the depth of the vertex labeled with $I_v$. During execution, for any node $v$, value of $d_v$ grows from $0$ to $\Theta(\log n)$, while interval $I_v$ narrows from $[1,n]$ to $[i,i]$ for some integer $i\in [n]$.

As mentioned earlier, the crash-resilient algorithm proceeds in phases, and there are $3\cdot\lceil\log{n}\rceil$ phases in total. Within each phase, there are three rounds.

The first round serves as committee announcement. Specifically, in this round, every committee member broadcasts a notification message to all nodes via $n$ links, including other committee members.\footnote{Here we adopt the standard terminology of the message-passing model: in a complete network of size $n$, each node has $n$ links, each connecting to one of the $n$ nodes in the system.} This allows every node to know which committee members are active.

In the second round, every node will send a message to all active committee members to inform its current status.
Specifically, for each node $v$, it sends $\langle\id(v),I_v,d_v,p_v\rangle$. On the other hand, for every committee member $u$, after receiving the messages from other nodes, $p_u$ will be updated to be the maximum received $p_v$ value. This mechanism ensures the consistency of $p$ values among all nodes, which helps build a committee of appropriate size.
In the following, for the ease of presentation, we use $M_u$ to denote the set of messages that a committee member $u$ has received during round two of this phase.

The last round is the most complex one: if there are committee members who have not crashed, they attempt to halve the intervals and distribute their decision to all nodes. After receiving the decisions from the committee members, nodes will proceed as instructed by the committee members or elect a new committee if they realize all existing committee members have crashed.

Specifically, for each committee member $u$, in the third round, it only halves intervals whose corresponding depth equals to $\tilde{d}_u$, where $\tilde{d}_u$ is the minimum depth value among all messages in $M_u$. Working only on the minimum depth helps maintain consistency among all remaining nodes: any node $w$ currently having depth $d_w>\tilde{d}_u$ will not forward to depth $d_w+1$ until all other nodes reach depth $d_w$. The halving strategy itself is simple: for any node $w$ with $d_w=\tilde{d}_u$ and interval $I_w$, we compare its identifier $\id(w)$ with the nodes that currently also choose interval $I_w$; if the rank of $\id(w)$ is in the smaller half then we update $I_w$ to $\texttt{bot}(I_w)$, otherwise we update $I_w$ to $\texttt{top}(I_w)$. Here are the details of the committee members' actions.
For each message $\langle\id(w),I_w,d_w,p_w\rangle \in M_u$ with $d_w>\tilde{d}_u$, committee member $u$ simply responses $\langle\id(w),I_w,d_w,p_u\rangle$, replacing $p_w$ with $p_u$.
For each message $\langle\id(w),I_w,d_w,p_w\rangle \in M_u$ with $d_w=\tilde{d}_u$, committee member $u$ will first filter out all $\langle\id(v),I_v,d_v,p_v\rangle$ messages in $M_u$ satisfying $I_v=I_w$. Let $ID_{(u,w)}$ be the set containing all identities in the filtered messages, notice that $\id(w)\in ID_{(u,w)}$. Then, node $u$ filters out all $\langle\id(v),I_v,d_v,p_v\rangle$ messages in $M_u$ satisfying $I_v\subseteq  \texttt{bot}(I_w)$. Let $B_{(u,w)}$ be the set containing all identities in the filtered messages. If $|B_{(u,w)}|+\texttt{rank}(\id(w), ID_{(u,w)}) \leq |\texttt{bot}(I_w)|$, then committee member $u$ will send response $\langle\id(w),\texttt{bot}(I_w),d_w+1,p_u\rangle$ to node $w$. Otherwise, $u$ will send response $\langle\id(w),\texttt{top}(I_w),d_w+1,p_u\rangle$ to node $w$. Here, $\texttt{rank}(\id(w), ID_{(u,w)})$ returns the rank of $\id(w)$ in set $ID_{(u,w)}$. This mechanism ensures potentially different responses from the committee members (caused by crashes) will not result in conflicts in nodes' new identities.

Once committee members have sent out their responses, nodes will act accordingly or elect a new committee if they realize all committee members have crashed.
More specifically, for a node $v$, if it has not received any message throughout the third round, then it assumes all committee members have crashed and updates $p_v$ to $p_v+1$, it will also elect itself as a committee member with probability $(256\cdot2^{p_v}\cdot\log{n})/n$.
Otherwise, if $v$ receives at least one message during the third round, then it will update its $I_v, p_v, d_v$ accordingly as follows.
Let $R_v$ denote the messages $v$ received from the committee.
Sort the messages in $R_v$ by $d$ value in decreasing order.
Then, for messages with identical $d$ value, sort them by the left endpoint of $I$ in increasing order.
After sorting, assume $\langle\id_1, I_1,d_1,p_1\rangle$ is the first message in $R_v$. If $|I_v|>1$, node $v$ updates $d_v$ to $d_1$ and $I_v$ to $I_1$. (That is, $v$ updates $d_v$ and $I_v$ if its identity is not determined yet.) Let $\hat{p}_v$ denote the maximum $p$ value in $R_v$. If $\hat{p}_v >p_v$, then node $v$ updates $p_v$ to $\hat{p}_v$ and elects itself as a committee member with probability $(256\cdot2^{p_v}\cdot\log{n})/n$. The committee election mechanism in round three guarantees that when all existing committee members crash, the probability for the remaining nodes to become new committee members doubles. As a result, as the algorithm proceeds, the adversary needs to crash more and more nodes to stop the algorithm from succeeding, implying the algorithm's complexity scales with the actual number of failures.

Complete pseudocode of the algorithm is provided in Appendix~\ref{secapp:pseudocode-crash}.

\subsection{Overview of Analysis}

In this section, we sketch the analysis for the crash-resilient algorithm, see Appendix~\ref{secapp:analysis-crash} for complete proofs. Before diving into the details, we introduce some notations to facilitate presentation.

\begin{definition}\label{def:crash-analysis}
For any $0\leq k\leq 3\cdot\lceil\log{n}\rceil$ and any $1\leq j\leq 3$, at the end of round $j$ in phase $k$:
\begin{itemize}[nosep,itemsep=1pt]
	\item For any active node $v$ that has not determined its new identity (i.e., $|I_v|>1$), let $d_{k,j}(v)$ denote the value of $d_v$. (For convenience, let $d_{0,j}(v)$ denote the value of $d_v$ after initialization.)
	\item For any active node $v$, let $p_{k,j}(v)$ and $I_{k,j}(v)$ denote the value of $p_v$ and $I_v$, respectively. (For convenience, let $p_{0,j}(v)$ and $I_{0,j}(v)$ denote the value of $p_v$ and $I_v$ after initialization, respectively.)
	%\item For any active node $v$, define $V_{k,j}(I_{k,j}(v)) = \{u~|~u\text{ is active and }I_{k,j}(u)\subseteq I_{k,j}(v)\}$.
	\item For any active node $v$, for any round $j'$ in phase $k'$ that equals to or prior to round $j$ in phase $k$, define $V_{k',j'}(I_{k,j}(v)) = \{u~|~u\text{ is active at the end of round }j'\text{ in phase }k'\text{ and }I_{k',j'}(u)\subseteq I_{k,j}(v)\}$.
	\item Define $\hat{d}_{k,j}=\max_{v} d_{k,j}(v)$, $\hat{p}_{k,j}=\max_{v} p_{k,j}(v)$; and $\tilde{d}_{k,j}=\min_{v} d_{k,j}(v)$, $\tilde{p}_{k,j}=\min_{v} p_{k,j}(v)$.
\end{itemize}
\end{definition}

\paragraph{Correctness.} We begin the analysis by arguing correctness: our algorithm ensures all correct nodes will eventually obtain non-colliding new identities.
To this end, our first technical lemma focuses on the scenario where at least one committee member survives one phase. In such case, after that phase, for the nodes that have not determined their identities, the minimum $d_v$ value among these nodes increases by at least one, implying the maximum interval size among these nodes is at least halved.

\begin{restatable}{lemma}{LemNoCrashIncreasingHeight}\label{lem:no-crash-increasing-height}
For any phase $k$ where $1\leq k\leq 3\cdot\lceil\log{n}\rceil$, if $\tilde{d}_{k-1,3}\leq \lceil \log n\rceil$ and some node $v$ has $elected_v$ equal to $true$ at the beginning of phase $k$ and does not crash throughout phase $k$, then $\tilde{d}_{k,3}\geq\tilde{d}_{k-1,3}+1$.
\end{restatable}

%Lemma~\ref{lem:no-crash-increasing-height} above shows that so long as some committee member exists, every node's interval size will continue to decrease, allowing it to eventually pick a new identity once its interval size is one.
The following lemma shows that by the end of each phase, for any interval $I_v$ chosen by some node $v$, the total number of nodes whose chosen intervals are subsets of $I_v$ will not exceed the size of $I_v$. Therefore, once every node's interval is of size one, they must have obtained distinct new identities.

\begin{restatable}{lemma}{LemProcessorLessInterval}\label{lem:processor-less-interval}
For any node $v$, at the end of any round $j$ (where $1\leq j\leq 3$) in any phase $k$ (where $1\leq k\leq 3\cdot\lceil\log{n}\rceil$), if $v$ has not crashed, then $|V_{k,j}(I_{k,j}(v))|\leq |I_{k,j}(v)|$.
\end{restatable}

\paragraph{Complexity.} In this part, we show several key lemmas regarding the value of $p_v$, as it implicitly affects the time and the message complexity of our algorithm.

First, if all committee members crash in phase $k$, then all remaining nodes will double their probability of becoming committee members in phase $k+1$. Though higher $p_v$ value results in larger committee and hence more messages, it also forces the adversary to crash more nodes to stop the algorithm from succeeding.

\begin{restatable}{lemma}{LemCrashRebuildCommittee}\label{lem:crash-rebuild-committee}
For any phase $k$ where $1\leq k\leq 3\cdot\lceil\log{n}\rceil$, by the end of phase $k$, if there does not exist any active node $v$ that has $elected_v$ equal to $true$, then $\tilde{p}_{k+1,3}\geq\tilde{p}_{k,3}+1$.
\end{restatable}

Another way for the adversary to disrupt algorithm execution is to manipulate the exact moments that committee members crash so that different nodes may have different $p_v$ values.
%(For example, a committee member may have successfully sent a response to some node $w$ before crashing; thus, $w$ keeps its $p_w$ value unchanged while every other node increases its $p_v$ value.)
Such difference---though inevitable---is bounded throughout our algorithm's execution. Particularly, by the end of each phase, the gap between the maximum and the minimum $p_v$ value is at most one.

\begin{restatable}{lemma}{LemBoundedDiffK}\label{lem:bounded-difference-k}
For any phase $k$ where $0\leq k\leq 3\cdot\lceil\log{n}\rceil$, it holds that $\hat{p}_{k,3}\leq\tilde{p}_{k,3}+1$.
\end{restatable}

The last two lemmas establish the connection between value of $p_v$, size of committee, and number of crashed nodes. Note that they hold probabilistically, while all previous lemmas hold deterministically.

\begin{restatable}{lemma}{LemBoundedCimmitteeUpper}\label{lem:bounded-committee-upper}
By the end of any phase $k$ where $1\leq k\leq 3\cdot\lceil\log{n}\rceil$, the number of nodes that \highlightblue{ever have $elected_v$ equal to $true$ (i.e., including the ones that have crashed by then)} is at most $O(\min\{2^{\hat{p}_{k,3}}\cdot\log{n},n\})$, with probability at least $1-n^{-3}$.
\end{restatable}

\begin{restatable}{lemma}{LemBoundedCimmitteeLower}\label{lem:bounded-committee-lower}
By the end of any phase $k$ where $3\leq k\leq 3\cdot\lceil\log{n}\rceil$, if $\hat{p}_{k,3}\geq 3$, then the adversary has crashed at least $\Omega(\min\{2^{\hat{p}_{k,3}}\cdot\log{n},n\})$ nodes, with probability at least $1-n^{-3}$.
\end{restatable}

\paragraph{Proof of main theorem.} We conclude this section with a proof of \Cref{thm:main-crash}.

We begin with time complexity.
For every two phases, by Lemma~\ref{lem:no-crash-increasing-height} and Lemma~\ref{lem:crash-rebuild-committee}, either the minimum $d_v$ value among all remaining nodes have increased by at least one, or the minimum $p_v$ value among all remaining nodes have increased by at least one.
Formally, for any $0\leq k\leq2\cdot\lceil\log{n}\rceil-2$, either $\tilde{d}_{k+2,3}\geq\tilde{d}_{k,3}+1$ or $\tilde{p}_{k+2,3}\geq\tilde{p}_{k,3}+1$.
Thus, after $2\cdot\lceil\log{n}\rceil$ phases, either $\tilde{d}_{2\cdot\lceil\log{n}\rceil,3} \geq  \lceil \log n\rceil$ or $\tilde{p}_{2\cdot\lceil\log{n}\rceil,3} \geq \lceil \log n\rceil$.
In the former case, every remaining node must have an interval of size one, hence obtaining a new identity in $[n]$. Moreover, Lemma~\ref{lem:processor-less-interval} ensures nodes' new identities are distinct.
In the latter case, by the end of phase $2\cdot\lceil\log{n}\rceil$, every remaining node $v$ has $p_{2\cdot\lceil\log{n}\rceil,3}(v)\geq\lceil\log{n}\rceil$. By algorithm description, in such case, every remaining node is a committee member by the end of phase $2\cdot\lceil\log{n}\rceil$.
%(Specifically, assume $v$ is not a committee member by then, let us consider the round $v$ changes its $p_v$ value to $\lceil\log{n}\rceil$. By algorithm description, this can only happen due to Line~\ref{line:node-change-pv-1} or Line~\ref{line:node-change-pv-2} of Figure~\ref{fig:alg-rename-node-action}. But in either case, node $v$ will set $elected_v$ to $true$ with probability one, implying it will join the committee, a contradiction.)
Note that the adversary can not crash all nodes. Hence, by Lemma~\ref{lem:no-crash-increasing-height}, after another $\lceil\log{n}\rceil$ phases, we must have $\tilde{d}_{3\cdot\lceil\log{n}\rceil,3}\geq\lceil\log{n}\rceil$, implying every node has obtained a unique identity in $[n]$ by then.

Next, we turn attention to message complexity. By Lemma~\ref{lem:bounded-committee-upper}, we know the number of nodes that ever have set $elected_v$ to $true$---i.e., the total number of committee members throughout the entire execution---is bounded by $O(\min\{2^{\hat{p}_{3\cdot\lceil\log n\rceil,3}}\cdot\log{n},n\})$ with probability at least $1-n^{-3}$. Note that in each of the $O(\log{n})$ rounds throughout algorithm execution, every active committee member exchanges messages with all nodes. Hence, the total number of messages sent by all nodes is bounded by $O(\min\{2^{\hat{p}_{3\cdot\lceil\log n\rceil,3}}\cdot\log{n},n\}\cdot n\log{n})$ with probability at least $1-n^{-3}$. Now, in case $\hat{p}_{3\cdot\lceil\log{n}\rceil,3}\leq2$, then the total message complexity is $O(n\log^2{n})$ with probability at least $1-n^{-3}$. Otherwise, in case $\hat{p}_{3\cdot\lceil\log{n}\rceil,3}\geq3$, by Lemma~\ref{lem:bounded-committee-lower}, the adversary has crashed at least $\Omega(\min\{2^{\hat{p}_{3\cdot\lceil\log n\rceil,3}}\cdot\log{n},n\})$ nodes with probability at least $1-n^{-3}$. Let $f$ denote the number of nodes crashed by the adversary, we conclude the message complexity of the algorithm is $O((f+\log{n})\cdot n\log{n})$, with probability at least $1-2\cdot n^{-3}$.

Finally, by algorithm description, clearly each message is of $O(\log{N})$ bits. Moreover, the algorithm never sends more that $\Theta(n^2\log{n})$ messages deterministically, as the algorithm has $O(\log{n})$ time complexity deterministically and the committee size is at most $n$ deterministically.

\section{Byzantine-resilient Renaming Algorithm}

\subsection{Algorithm Description}

Our Byzantine-resilient renaming algorithm consists of three parts: committee election, reaching consensus on nodes' new identities within committee, and distributing new identities.

\paragraph{Committee election and original identity aggregation.}

With the help of shared randomness, the committee election process becomes much simpler when compared with the crash-resilient algorithm. Specifically, when execution starts, each node creates a ``pool of committee candidates'' as follows: for each identity in the original namespace (that is, each value in $[N]$), that identity becomes a candidate with some predefined probability $p_0$. Since nodes use shared randomness to create this pool, all correct nodes have identical view on this pool. Then, if a node finds its identity is in this pool, it will notify all nodes to signal that it has been elected as a committee member. Note that the adversary may manipulate Byzantine nodes' behavior so that honest nodes may have different local views of the committee, but our algorithm ensures that these views mostly overlap and correct members always take the majority.

Once a committee is formed, another round is used for the committee members to aggregate the identities of all nodes. Specifically, each node $u$ sends its original identity $\id(u)\in[N]$ to the committee members in its view. By the end of this round, each committee member $v$ has obtained an \emph{identity list} $L_v$ which is a bit vector consisting of $N$ bits: $L_v[\id(u)] = 1$ iff the identity $\id(u)$ of node $u$ has been received by $v$.

\paragraph{Renaming within committee using fingerprinting.}

The second part of our algorithm, which is also the most interesting and the most complicated part, allows committee members to reach consensus on their identity lists efficiently and competitively. At a high level, this consensus process is performed recursively. Initially, the committee members attempt to agree on the entire list. If consensus is achieved, the process concludes. Otherwise, the committee recursively partitions the list into two segments and attempts to reach consensus on each part independently. This process continues until consensus is achieved for every segment. By then, for each identity in the list, its new identity is determined by its position in the list.

Throughout execution, for each committee member $v$, it uses a stack $J_v$ to store the segments that pend processing and uses a set $\hat{J}_v$ to store the segments that have already been processed. Specifically, each element in $J_v$ (and $\hat{J}_v$) is an interval $[l,r]$ corresponding to segment $L_v[l,\cdots,r]$. Initially, $J_v$ contains only $[1,N]$ and $\hat{J}_v$ is empty. Moreover, our algorithm ensures that at any moment, for any committee member $v$, $J_v$ and $\hat{J}_v$ together correspond to a partition of $[1,N]$, and all committee members share identical $J$ and $\hat{J}$.

We now describe the consensus process for each segment in detail. Consider an interval $[l,r]$ in $J$, which correspond to segment $L[l,\cdots,r]$. If the segment contains a single bit (i.e., $l=r$), then the committee members run a classical binary consensus procedure called \consensus\xspace, move interval $[l,r]$ to 
$\hat{J}$, and update the bit according to the consensus process's result. Otherwise, if $l<r$, then each committee member $v$ hashes $L_v[l,\cdots,r]$ with a hash function constructed via shared randomness, obtaining a fingerprint $s_v$ of size $O(\log{N})$ bits. Additionally, each committee member $v$ also keeps track of $cnt_v$, which counts the number of ones in $L_v[l,\cdots,r]$ (i.e., the number of identities appearing in $L_v[l,\cdots,r]$). Committee members need to agree on the tuple $\langle s_v, cnt_v\rangle$. Note that this tuple contains multiple bits, and running \consensus\xspace for each bit is inappropriate, as that would consume too much time, and the output does not necessarily preserve the semantics of the input. Instead, we leverage a procedure called \validator\xspace that produces $\langle same_v, \langle s'_v, cnt'_v\rangle \rangle$ as output, where $same_v$ is a single bit. This procedure enforces: (1) strong validity, in that $\langle s'_v, cnt'_v\rangle$ must be some correct node's input; and (2) weak agreement, in that if $same_v=1$ then $\langle s'_v, cnt'_v\rangle=\langle s'_u, cnt'_u\rangle$ for every correct committee member $u$. Once \validator\xspace is executed, committee members will run \consensus\xspace again to reach agreement on binary variable $same$. For each committee member $v$, denote the output of this execution of \consensus\xspace as $same'_v$.

At this point, for each committee member $v$, if $same'_v=1$, then it can be assured that every other correct committee member $u$ also has $same'_u=1$ and $\langle s'_v, cnt'_v\rangle=\langle s'_u, cnt'_u\rangle$. Unfortunately, this is insufficient to conclude the consensus process for segment $L[l,\cdots,r]$. In particular, $s'_v$ does not necessarily equal to the hash of $L_v[l,\cdots,r]$, and $v$ cannot recover the preimage of $s'_v$ due to the one-way property of hash functions. That is, reaching agreement on hash value does not imply reaching agreement on segment's original value. Moreover, it might be the case that the preimage of the agreed hash value is only known by a few committee members (i.e., only a few committee members satisfy $s_v=s'_v$). This creates difficulty for new identity distribution. To resolve this issue, for each committee member $v$, after obtaining $s'_v$, it sets a binary variable $diff_v$ to $1$ iff $s_v\neq s'_v$, and then sends $diff_v$ to other committee members. Then, after receiving $diff$ from other committee members, $v$ sets a binary variable $diff'_v$ to $1$ if it finds many other committee members $u$ report $diff_u=1$, otherwise $v$ sets $diff'_v$ to $diff_v$. Finally, each committee member $v$ runs \consensus\xspace to reach agreement on $diff'_v$, and we denote the output as $diff''_v$.

We can now conclude, the consensus process for segment $L[l,\cdots,r]$ is successful if the following two conditions are met: (1) every committee member $v$ agrees on the same hash value ($same_v' = 1$), and (2) every committee member $v$ finds committee members with appropriate original segment form the majority ($diff_v'' = 0$). If the consensus process for segment $L[l,\cdots,r]$ fails, then the interval $[l,r]$ is divided into two halves, and the algorithm recurse into each of them. Otherwise, if the consensus process succeeds, then there are two possible scenarios for a committee member $v$. On the one hand, if $v$'s segment matches the agreed hash value (i.e., $\langle s_v, cnt_v\rangle=\langle s'_v, cnt'_v\rangle$), then it will participate in the process of distributing new identities to the nodes within $L_v[l,\cdots,r]$ later. On the other hand, if $v$'s segment does not match the agreed hash value, then it will mark the interval $[l,r]$ as \emph{dirty}, and replace $L_v[l,\cdots,r]$ with an arbitrary binary string that contains exactly $cnt_v'$ $1$s. Also, later $v$ will not participate in the process of distributing new identities to the nodes within $L_v[l,\cdots,r]$. This mechanism ensures a committee member's dirty intervals will not affect its ability to correctly distributed new identities for nodes within non-dirty intervals.

\paragraph{Distribute new identities.}

After committee members reach consensus on every segment of the identity list, they distribute new identities. Specifically, when committee member $v$ distributes a new identity to node $u$, it first checks whether $u$'s original identity is within an interval marked as dirty. If not, then $v$ sends $\langle NEW, \nid(u)\rangle$ to $u$, where $\nid(u)$ equals the rank of $u$'s original identity in $L_v$. (Here, ``rank'' means the number of $1$s in $L_v$ that occur before position $\id(u)$.) Otherwise, $v$ simply sends $\langle NEW,\texttt{null}\rangle$ to $u$.
On the other hand, for each node, it waits for sufficiently many $NEW$ messages from the committee, and sets its new identifier as the majority value among these $NEW$ messages.

Complete pseudocode of the algorithm is provided in Appendix~\ref{secapp:pseudocode-byz}.

\subsection{Discussion}\label{subsec:byz-discussion}

Our Byzantine-resilient renaming algorithm achieves strong fault-tolerance with low communication cost and good scalability. In exchange for these favorable properties, it relies on several assumptions. In this part, we discuss the feasibility of these assumptions and how some of them could be weaken or even removed.

Shared randomness plays two roles in our algorithm. First, it helps correct nodes reach agreement on the constitution of the committee. Second, it helps committee members construct hash functions.
We suspect this assumption could be removed at the cost of a more complicated algorithm.
Specifically, electing and maintaining a small committee consisting of mostly correct nodes with limited cost is possible, but highly non-trivial, see \cite{augustine20} for an example.
Once a committee is formed, one can leverage existing techniques to generate shared randomness (see, e.g., \cite{cachin00,shoup00}), hence construct hash functions.

Message authentication, on the other hand, prevents Byzantine nodes from masquerading as honest or other non-existent nodes. Many works on Byzantine fault-tolerance also make similar assumption, such as \cite{dolev83,fitzi09}. In practice, message authentication is often achieved by using digital signature, which in turn can be achieved by a public key infrastructure (PKI). PKI has been widely used in practice nowadays. We also stress that, the availability of PKI does \emph{not} mean every node in a distributed system knows the public keys of all nodes (this would render renaming trivial). Instead, in practical systems, every node only needs to know a constant number of public keys (\emph{none} of which correspond to a node in the system), and verifying another node's signature is done via a mechanism known as certificate chain.

The assumption that the adversary is non-adaptive seems critical for the committee based approach. Specifically, an adaptive adversary can start acting maliciously after the committee has been elected, violating the key property that most of the committee members are correct. Therefore, how to achieve Byzantine-resilient renaming against an adaptive adversary with low communication cost is an interesting direction worth further exploration.

\subsection{Overview of Analysis}

In this subsection, we provide an overview for the analysis of the Byzantine-resilient renaming algorithm. See Appendix~\ref{secapp:analysis-byz} for complete proofs. Throughout, we assume there are at most $(1/3-\epsilon_0)n$ Byzantine nodes, where $\epsilon_0>0$ is an arbitrarily small constant. We further define $\highlight{p_0=8\log{n}/((1-3\epsilon_0)\epsilon_0^2n)}$, $\highlight{c_g=(1-3\epsilon_0/2)(2/3+\epsilon_0)p_0n}$, and $\highlight{\hat{c}_g=4p_0n}$.
To simplify notation, for a length-$N$ bit vector $L\in\{0,1\}^{N}$ and an interval $j=[l,r]$, where $l,r\in [N]$ and $l\leq r$, we use $L[j]$ to denote segment $L[l,\cdots,r]$.

Due to the presence of Byzantine nodes, honest nodes may have different views on the set of nodes that are present in the system, as well as the members in the committee. To facilitate presentation, we further define following notations.

\begin{definition}\label{def:notations-byz}
Let $V$ be the set of all nodes. Denote $G\subseteq V$ as the set of correct committee members. For each node $v\in V$, denote $C_v\subseteq V$ as the set of committee members that are in $v$'s view. Denote $B\subseteq V$ as the set of Byzantine nodes in $\bigcup_{v\in V} C_v$.
\end{definition}

\paragraph{Preliminary.}

We first introduce three auxiliary tools used in our algorithm, and we begin with hash functions that can guarantee small collision probability.

\begin{fact}[Hash function~\cite{brody2014beyond}]\label{fact:hash}
For any set $S\subseteq[U]$ of size $|S|\geq 2$ and $i>0$, a random hash function $\hash:[U]\to [T]$ where $T\in O(|S|^{i+2})$ has no collisions with probability at least $1-1/|S|^{i}$. That is, for all $x,y\in S$ with $x\neq y$, we have $\hash(x)\neq \hash(y)$, with probability at least $1-1/|S|^{i}$. Moreover, a random hash function satisfying such a guarantee can be constructed using $O(\log{U})$ random bits.
\end{fact}

Next, we introduce a procedure called \validator\xspace inspired by Lenzen and Sheikholeslami~\cite{lenzen22}. Compared with classical consensus, it provides stronger validity property but weaker agreement property.

\begin{restatable}[Validator]{lemma}{LemValidator}\label{lem:validator}
Suppose each node $v\in G$ receives input $in_v\in\{0,1\}^{k}$ where $k\in O(\log{N})$. Assume $G\subseteq\bigcap_{v\in G} C_v$ and $c_g\leq |G|\leq \hat{c}_g$ and $|B|<c_g/2$, there exists a $2$-round algorithm \validator\xspace with message complexity $O(\hat{c}_g^2)$ and bit complexity $O(\hat{c}_g^2\log{N})$ that produces output $\langle same_v,out_v\rangle$ for each $v\in G$, where $same_v\in \{0,1\}$ and $out_v\in \{0,1\}^k$, satisfying:
\begin{itemize}[nosep,itemsep=1pt]
	\item validity property: (1) $out_v=in_u$ for some $u\in G$; moreover, (2) if $in_u=in\in\{0,1\}^k$ for all $u\in G$, then $same_v=1$ and $out_v=in$.
	\item weak agreement property: if $same_v=1$, then $out_u=out_v$ for all $u\in G$.
\end{itemize}
\end{restatable}

Lastly, we also utilize classical binary consensus in our algorithm.

\begin{restatable}[Consensus]{lemma}{LemConsensus}\label{lem:consensus}
Suppose each node $v\in G$ receives input $in_v\in\{0,1\}$. Assume $G\subseteq\bigcap_{v\in G} C_v$ and $c_g\leq |G|\leq \hat{c}_g$ and $|B|<c_g/2$, there exists an $O(\hat{c}_g)$-round algorithm \consensus\xspace with message complexity $O(\hat{c}_g^3)$ and bit complexity $O(\hat{c}_g^3\log{N})$ that produces $out_v\in\{0,1\}$ for each $v\in G$ satisfying:
\begin{itemize}[nosep,itemsep=1pt]
	\item validity property: $out_v=in_u$ for some $u\in G$.
	\item agreement property: $out_v=out_u$ for all $u\in G$.
\end{itemize}
\end{restatable}

\paragraph{Analysis of main algorithm.} We now proceed to analyze our algorithm. The first lemma states that for each correct node, the committee in its view contains all correct committee members and limited number of Byzantine nodes. This is achieved with the help of shared randomness and message authentication.

\begin{restatable}{lemma}{LemCommitteeAnnouncement}\label{lem:committee-announcement}
For each correct node $v$, denote $C_v$ as the committee members in $v$'s view. Then, $G\subseteq C_v$. Moreover, with probability at least \highlight{$1-n^{-5}$}, it holds that $c_g\leq|G|\leq\hat{c}_g$ and $|B|<c_g/2$.
\end{restatable}

The following observation is immediate by algorithm description and the assumption that messages are authenticated (hence Byzantine nodes cannot fake their identities.)

\begin{fact}\label{fac:announce}
After nodes have sent their original identities to the committee, the following holds for any $i\in [N]$.
\begin{enumerate}[nosep,itemsep=1pt]
	\item\label{item:announce-1} If $i$ is the identifier of a correct node, then $L_v[i]=1$ for any correct committee member $v$.
	\item\label{item:announce-2} If $L_v[i]=1$ for some correct committee member $v$, then $i$ must be the identifier of a node.
\end{enumerate}
\end{fact}

In the remaining analysis, to simplify presentation, we assume the following properties hold. Later when we wrap-up, we argue they hold with high probability throughout the execution.

\begin{property}\label{cond:byz}
\begin{enumerate}
	\item\label{item:cond-byz-1} $c_g\leq|G|\leq\hat{c}_g$ and $|B|<c_g/2$.
	\item\label{item:cond-byz-3} For any two correct committee members $u,v$, for any two intervals $j_u,j_v\subseteq[N]$ of same length, if $L_u[j_u]\neq L_v[j_v]$, then $\hash(L_u[j_u])\neq\hash(L_v[j_v])$.
\end{enumerate}
\end{property}

Assuming \Cref{cond:byz}, the following lemma states that all correct committee members partition and process segments in their identity lists in the same manner.

\begin{restatable}{lemma}{LemSameJv}\label{lem:while-same-Jv}
Assume \Cref{item:cond-byz-1} of \Cref{cond:byz} holds, throughout algorithm execution, all correct committee members $v$ have identical $J_v$ and $\hat{J}_v$; moreover, $J_v\cup\hat{J}_v$ is a partition of $[N]$.
\end{restatable}

Next, we show that for each segment in the identity list, if that segment is of size one, or if the value of that segment is identical for all correct committee members, then that segment will not be partitioned further. This is enforced by the validity property of \consensus\xspace and \validator.

\begin{restatable}{lemma}{LemSamejv}\label{lem:while-same-jv}
Assume \Cref{item:cond-byz-1} of \Cref{cond:byz} holds, in any iteration of the \textbf{while} loop (Line~\ref{line:byz-alg-while-loop} in \Cref{fig:alg-rename-byz} in Appendix~\ref{secapp:pseudocode-byz}), all correct committee members $v$ work on the same interval $j_v$. Moreover, if all correct committee members have identical $L_v[j_v]$ or $|j_v|=1$, then all correct committee members add $j_v$ to $\hat{J}_v$.
\end{restatable}

With above lemma, we can bound the runtime of our algorithm. Note that the runtime scales with the \emph{actual} number of Byzantine nodes, not the \emph{upper bound} on the number of Byzantine nodes.

\begin{restatable}{lemma}{LemWhileLoopCount}\label{lem:while-loop-iterations}
Assume \Cref{item:cond-byz-1} of \Cref{cond:byz} holds, the \textbf{while} loop terminates within $4f\log{N}$ iterations.
\end{restatable}

The following lemma suggests that, for each segment of the identity list, our algorithm not only ensures \emph{all} correct committee members reach agreement on the \emph{hash} of this segment and the \emph{number of identities} in this segment; moreover, \emph{most} correct committee members reach agreement on the \emph{value} of this segments. This property guarantees the committee can correctly distribute new identities later.

\begin{restatable}{lemma}{LemMajorityCommitteeConsensus}\label{lem:majority-committee-consensus}
Assume both items in \Cref{cond:byz} hold, at the end of each iteration of the \textbf{while} loop, for any correct committee member $v$, for any interval $j=[l,r]\in \hat{J}_v$, it holds that:
\begin{enumerate}[nosep,itemsep=1pt]
	\item\label{item:majority-1} There exist at least $c_g/2$ correct committee members $u$ satisfying: (a) $u$ does not mark interval $j$ as dirty, (b) $L_u[j]=L_v[j]$, and (c) if $i\in[l,r]$ is the identifier of a correct node then $L_u[i]=1$.
	\item\label{item:majority-2} All correct committee members reach consensus on the number of $1$s in $L_v[j]$, which is at most \highlight{the number of nodes with original identities} in interval $j$.
\end{enumerate}
\end{restatable}

\Cref{lem:while-same-Jv} and above lemma together imply the following corollary, which states that our algorithm can correctly distribute new identities.

\begin{restatable}{lemma}{LemByzIDUnique}\label{lem:byz-id-unique}
Assume both items in \Cref{cond:byz} hold, every correct node gets its new unique identifier in $[n]$. Moreover, this renaming is order-preserving.
\end{restatable}

\paragraph{Proof of main theorem.}

We conclude with a proof for \Cref{thm:main-byz}.

To begin with, we show \Cref{cond:byz} holds throughout execution with high probability.
For \Cref{item:cond-byz-1} of \Cref{cond:byz}, by \Cref{lem:committee-announcement}, it holds with probability at least \highlight{$1-n^{-5}$}.
Now, assuming \Cref{item:cond-byz-1} of \Cref{cond:byz}, there are at most $\hat{c}_g$ correct committee members. By \Cref{fact:hash}, in each iteration of the \textbf{while} loop, the probability that there is a collision when correct committee members invoke \texttt{Hash} is at most $(4p_0n)^2/N^6$. Recall \Cref{lem:while-loop-iterations}, throughout the entire execution, the probability that there is a collision is at most $4f\log N\cdot (4p_0n)^2/N^6\leq n^{-4}$.
Hence, assuming \Cref{item:cond-byz-1} of \Cref{cond:byz}, \Cref{item:cond-byz-3} of \Cref{cond:byz} holds with probability at least \highlight{$1-n^{-4}$}.
As a result, \Cref{cond:byz} holds with high probability in $n$.

The correctness of our algorithm is immediate by \Cref{lem:byz-id-unique}.

As for time complexity, by \Cref{lem:while-loop-iterations}, the \textbf{while} loop of the algorithm terminates in $O(f\cdot \log N)$ iterations. The time complexity of each iteration is dominated by running \validator{} and \consensus{}. Specifically, there are at most one execution of \validator{} and two executions of \consensus{} in each iteration. Each execution of \validator{} costs $2$ rounds, while each execution of \consensus{} costs $O(\log n)$ rounds.
Thus, the algorithm has a total round complexity of $O(f\cdot\log{N}\cdot\log{n})$.

Lastly, consider message complexity and bit complexity.
By algorithm description, committee announcement costs $O(n\log n)$ messages.
Similarly, original identity announcement costs $O(n\log n)$ messages.
By \Cref{lem:while-loop-iterations}, the \textbf{while} loop contains $O(f\log N)$ iterations.
Within each iteration, the message complexity is dominated by executions of \validator{} and \consensus{}.
Specifically, each iteration contains at most one execution of \validator{} and two executions of \consensus{}.
Each execution of \validator{} costs $O(\log^2 n)$ messages and each execution of \consensus{} costs $O(\log^3 n)$ messages.
Hence, the \textbf{while} loop in total costs \highlight{$O(f\cdot\log{N}\cdot\log^3{n})$} messages.
The new identity distribution process cost $O(n\log n)$ messages.
We conclude, the algorithm costs \highlight{$O(f\log{N}\log^3{n} + n\log{n})$} messages in total. Note that each message contains $O(\log N)$ bits. Therefore, the bit complexity of the entire algorithm is \highlight{$O((f\log{N}\log^3{n} + n\log{n})\log{N})$}.

%%% bib
\addcontentsline{toc}{section}{Bibliography}
\bibliographystyle{plain}
\bibliography{arxiv-full-v1.bib}

% % % appendix % % %

\clearpage
\appendix

\section*{Appendix}

\section{Pseudocode of the Crash-resilient Algorithm}\label{secapp:pseudocode-crash}

\begin{figure}[!h]
\hrule
\vspace{1ex}
\textbf{The crash-resilient renaming algorithm executed at each node $v$:}
\vspace{1ex}
\hrule
\begin{small}
\begin{algorithmic}[1]
\State $I_v\gets[1,n]$, $p_v\gets 0$, $d_v\gets 0$, $elected_v\gets false$.
\State Set $elected_v$ to $true$ with probability $(256\log{n})/n$. \Comment{End of initialization.}
\For {($j=1$ \textbf{to} $3 \cdot \lceil \log{n}\rceil $)}
	\If {($elected_v==true$)} \Comment{Start of round 1.}
		\State Send a notification to all nodes indicating $v$ is a committee member via $n$ links.
	\EndIf
	\For {(each link $i$ from $1$ to $n$ where a committee notification is received)} \Comment{Start of round 2.}
		\State Send message $\langle\id(v),I_v,d_v,p_v\rangle$ through link $i$. \label{line:node-send-msg}
	\EndFor
	\If {($elected_v==true$)}
		\State $M_v\gets\text{ all messages received in this round}$.
		\State $p_v\gets\text{ maximum }p\text{ value among all messages in }M_v$. \label{line:committee-change-pv}
	\EndIf
	\If {($elected_v==true$)} \Comment{Start of round 3.}
		\State Execute procedure \textbf{\texttt{CommitteeAction}}.\label{line:committee-decesion}
	\EndIf
	\State Execute procedure \textbf{\texttt{NodeAction}}.
\EndFor
\State Return the unique value in interval $I_v$ as new identity.
\end{algorithmic}
\end{small}
\hrule
\vspace{1ex}\caption{The crash-resilient renaming algorithm executed at each node.}\label{fig:alg-rename-main}\vspace{-4ex}
\end{figure}

\begin{figure}[!h]
\hrule
\vspace{1ex}
\textbf{Procedure \texttt{CommitteeAction} executed at each committee member $v$:}
\vspace{1ex}
\hrule
\begin{small}
\begin{algorithmic}[1]
\State $\tilde{d}_v \gets\text{ minimum }d\text{ value among all messages in }M_v$. \label{line:compute-tilde-d}
	\For {(every message $\langle\id(w),I_w,d_w,p_w\rangle\in M_v$ received through link $i_w$)}
		\If {($\tilde{d}_v==d_w$)}
			\State $ID_{(v,w)}\gets\{\id(u)~|~\langle\id(u),I_u,d_u,p_u\rangle\in M_v\text{ and }I_u==I_w\}$. \label{line:set-ID-def}
			\State $B_{(v,w)}\gets\{\id(u)~|~\langle\id(u),I_u,d_u,p_u\rangle\in M_v\text{ and }I_u\subseteq \texttt{bot}(I_w)\}$. \label{line:set-B-def}
			\If {($|B_{(v,w)}|+\texttt{rank}(\id(w),ID_{(v,w)})\leq|\texttt{bot}(I_w)|$)} %\Comment{$\texttt{rank}(e,S)$ determines the rank of element $e$ in set $S$.}
				\State Send response message $\langle\id(w),\texttt{bot}(I_w),d_w+1,p_v\rangle$ through link $i_w$.
			\Else
				\State Send response message $\langle\id(w),\texttt{top}(I_w),d_w+1,p_v\rangle$ through link $i_w$.
			\EndIf
		\Else
			\State Send response message $\langle\id(w),I_w,d_w,p_v\rangle$ through link $i_w$.
		\EndIf
	\EndFor
\end{algorithmic}
\end{small}
\hrule
\vspace{1ex}\caption{Procedure \texttt{CommitteeAction} executed at each committee member.}\label{fig:alg-rename-committee-action}\vspace{-4ex}
\end{figure}

\begin{figure}[!h]
\hrule
\vspace{1ex}
\textbf{Procedure \texttt{NodeAction} executed at each node $v$:}
\vspace{1ex}
\hrule
\begin{small}
\begin{algorithmic}[1]
\If {(no message is received in this phase)} %\Comment{All committee members have crashed.}
	\State $p_v\gets p_v+1$.\label{line:no-message-received-from-committee} \label{line:node-change-pv-1}
	\State Set $elected_v$ to $true$ with probability $(\highlightblue{256}\cdot2^{p_v}\cdot\log{n})/n$.
\Else %\Comment{One or multiple response messages are received.}
	\State $R_v\gets\text{ all response messages from committee members}$.
	\State Sort $R_v$ first by $d$ value in decreasing order, then by $\texttt{min}(I)$ in increasing order.
	\State Let $m_1=\langle\id_1, I_1, d_1, p_1\rangle$ be the first message in $R_v$ after sorting.
	\If {($|I_v|>1$)}
		$d_v\gets d_1$, $I_v\gets I_1$. \label{line:node-change-d-and-interval}
	\EndIf
	\State $\hat{p}_v\gets\text{ the maximum }p\text{ value in }R_v$.
	\If {($\hat{p}_v>p_v$)}
		\State $p_v\gets\hat{p}_v$. \label{line:node-change-pv-2}
		\If {($elected_v==false$)}
			Set $elected_v$ to $true$ with probability $(256\cdot2^{p_v}\cdot\log{n})/n$.
		\EndIf
	\EndIf
\EndIf
\end{algorithmic}
\end{small}
\hrule
\vspace{1ex}\caption{Procedure \texttt{NodeAction} executed at each node.}\label{fig:alg-rename-node-action}\vspace{-4ex}
\end{figure}

\section{Proofs for the Analysis of the Crash-resilient Algorithm}\label{secapp:analysis-crash}

\LemNoCrashIncreasingHeight*
\begin{proof}
Suppose node $v$ has $elected_v=true$ at the beginning of phase $k$ and $v$ does not crash throughout phase $k$. It sends a notification to all nodes in the first round of phase $k$. According to Line~\ref{line:node-send-msg} of Figure~\ref{fig:alg-rename-main}, in the second round of phase $k$, every remaining node $w$ would include $d_{k-1,3}(w)$ as the $d$ value in the message for $v$.
%By definition, $d_{k-1,3}(w)\geq\tilde{d}_{k-1,3}$.
These messages together form the received message set of node $v$, which is denoted by $M_v$.

According to Figure~\ref{fig:alg-rename-committee-action}, after gathering $M_v$, committee member $v$ first computes $\tilde{d}_v$ to be the minimum $d$ value among all messages in $M_v$.
By algorithm description, it is easy to verify that $\tilde{d}_{k,2} \geq \tilde{d}_v\geq \tilde{d}_{k-1,3}$.
Now, for every message $\langle\id(w),I_w,d_w,p_w\rangle$ in $M_v$ with $d_w=\tilde{d}_v$, committee member $v$ will send a response containing a $d$ value that equals to $\tilde{d}_v+1$ to node $w$. On the other hand, for every message $\langle\id(w),I_w,d_w,p_w\rangle$ in $M_v$ with $d_w\neq\tilde{d}_v$, by the definition of $\tilde{d}_v$, it must be $d_w\geq\tilde{d}_v+1$. Hence, for each such message, committee member $v$ will send a response containing a $d$ value that equals $d_w$. Therefore, we can conclude, during round three of phase $k$, every remaining node must have received a response from $v$ containing a $d$ value that is at least $\tilde{d}_v+1$.

According to Figure~\ref{fig:alg-rename-node-action}, during round three of phase $k$, for a node $w$ that has not determined its identity, if $w$ receives any response from committee members, it will update its $d$ value to be the maximum $d$ value among all responses. As a result, by our above analysis, $d_{k,3}\geq\tilde{d}_v+1$. This further implies $\tilde{d}_{k,3}\geq\tilde{d}_v+1$. Recall that $\tilde{d}_v\geq\tilde{d}_{k-1,3}$, the lemma is proved.
\end{proof}

\LemProcessorLessInterval*
\begin{proof}
Recall \Cref{def:crash-analysis}, at the end of round $j$ in phase $k$, for any node $v$ that is still active, $V_{k,j}(I_{k,j}(v))$ contains active nodes whose chosen intervals are subsets of $I_v$ (i.e., $I_{k,j}(v)$) at that moment.

We prove the lemma by induction on rounds, and we begin by considering the basis, which is the case that $k=0$. Recall that during initialization, every node $v$ sets $I_v=[1,n]$. Thus, for every node $v$, for any $1\leq j\leq 3$, we have $|V_{0,j}(I_{0,j}(v))|\leq n = |I_{0,j}(v)|$.

Assuming the lemma is true up to the end of phase $k$. For the inductive step, consider the three rounds in phase $k+1$. Fix a node $v$ that does not crash throughout phase $k+1$.

We first consider the first and the second round in phase $k+1$. By algorithm description, $v$ only updates its interval $I_v$ in the third round in each phase, after receiving responses from committee members. Thus, $I_{k+1,1}(v)=I_{k+1,2}(v)=I_{k,3}(v)$, implying $V_{k+1,2}(I_{k+1,2}(v)) \subseteq V_{k+1,1}(I_{k+1,1}(v)) \subseteq V_{k,3}(I_{k,3}(v))$. On the other hand, by the induction hypothesis, $|V_{k,3}(I_{k,3}(v))|\leq|I_{k,3}(v)|$. As a result, we conclude
$$|V_{k+1,1}(I_{k+1,1}(v))|\leq|I_{k,3}(v)|=|I_{k+1,1}(v)|\text{ and }|V_{k+1,2}(I_{k+1,2}(v))|\leq|I_{k,3}(v)|=|I_{k+1,2}(v)|,$$
which completes the inductive step for the first two rounds of phase $k+1$.

Next, we consider the third round of phase $k+1$. If at the beginning of the third round in phase $k+1$ all committee members have crashed, then every remaining active node will not change its interval in the third round. In such case, by an analysis similar to above, we have $|V_{k+1,3}(I_{k+1,3}(v))|\leq|I_{k,3}(v)|=|I_{k+1,3}(v)|$. Another simple case is $I_{k+1,3}(v)=[1,n]$, in which $|V_{k+1,3}(I_{k+1,3}(v))|\leq|I_{k+1,3}(v)|$ trivially holds.

So in the remaining analysis, assume there are active committee members at the beginning of round three in phase $k+1$ and $I_{k+1,3}(v)\neq[1,n]$.
Let $\hat{I}_{k+1,3}(v)$ denote the interval with the largest phase index $k'<k+1$ satisfying $I_{k',3}(v)\neq I_{k+1,3}(v)$. (That is, $\hat{I}_{k+1,3}(v)$ is the value of $I_v$ prior to $I_{k+1,3}(v)$.) Clearly, either $\texttt{bot}(\hat{I}_{k+1,3}(v)) = I_{k+1,3}(v)$ or $\texttt{top}(\hat{I}_{k+1,3}(v)) = I_{k+1,3}(v)$. Depending on the value of $k'$, we consider two complement scenarios.

\smallskip\textsc{Scenario 1: $k'=k$.} This implies $v$ has halved its interval in the third round of phase $k+1$. We further consider the following two sub-scenarios.
\begin{itemize}
	\item Scenario 1.1: $I_{k+1,3}(v)=\texttt{bot}(I_{k,3}(v))$. Define $U$ as the set of nodes that have identical interval with $v$ at both the end of phase $k$ and phase $k+1$. That is,
	$$U=\{u\mid I_{k+1,3}(u)=I_{k+1,3}(v), I_{k,3}(u)=I_{k,3}(v)\}.$$
	Notice that $U\neq\emptyset$ as $v\in U$. Let $u$ be the node with the \emph{largest} identifier in $U$. Since $u$ halved its interval in phase $k+1$ (particularly, $I_{k+1,3}(u)=\texttt{bot}(I_{k,3}(u))$), during the second round in phase $k+1$, there must exist a committee member $c$ that finds $|B_{(c,u)}|+\texttt{rank}(\id(u),ID_{(c,u)})\leq|\texttt{bot}(I_{k,3}(u))|$. (Recall Line~\ref{line:set-ID-def} and Line~\ref{line:set-B-def} in \Cref{fig:alg-rename-committee-action} for the definition of $B_{(c,u)}$ and $ID_{(c,u)}$.)
	
	Now, consider node set $V_{k+1,3}(I_{k+1,3}(v))$, it can be partitioned into two subset: (1) the ones that do not halve their intervals in phase $k+1$, henceforth called $V^{unchanged}_{k+1,v}$; and (2) the ones that halve their intervals in phase $k+1$, henceforth called $V^{halved}_{k+1,v}$. That is,
	$$V^{unchanged}_{k+1,3}=V_{k+1,3}(I_{k+1,3}(v)) \cap V_{k,3}(I_{k+1,3}(v))\text{ and }V^{halved}_{k+1,3}=V_{k+1,3}(I_{k+1,3}(v)) \setminus V_{k,3}(I_{k+1,3}(v)).$$
	
	On the one hand, for any node $w\in V^{unchanged}_{k+1,3}$: $I_{k,3}(w)=I_{k+1,3}(w)=I_{k+1,3}(v)=\texttt{bot}(I_{k,3}(v))=\texttt{bot}(I_{k,3}(u))$. By the definition of $B_{(c,u)}$, we have $V^{unchanged}_{k+1,3} \subseteq \{w\mid\id(w)\in B_{(c,u)}\}$.
	
	On the other hand, for any node $w\in V^{halved}_{k+1,3}$, it must in set $U$. Since $u$ has the largest identifier in $U$, we have $\{\id(w)\mid w\in V^{halved}_{k+1,3}\} \subseteq \{i\mid i\in ID_{(c,u)}, i\leq\id(u)\}$.
	
	At this point, we conclude that
	$$|V_{k+1,3}(I_{k+1,3}(v))| \leq |B_{(c,u)}| + |\{i\mid i\in ID_{(c,u)}, i\leq\id(u)\}| \leq |\texttt{bot}(I_{k,3}(u))| = |I_{k+1,3}(v)|,$$
	which competes the inductive step for the third round of phase $k+1$ for Scenario 1.1.
	
	\item Scenario 1.2: $I_{k+1,3}(v)=\texttt{top}(I_{k,3}(v))$. Define $U$ in the same manner as in Scenario 1.1, let $u$ be the node with the \emph{smallest} identifier in $U$. Again, there must exist a committee member $c$ that finds $|B_{(c,u)}|+\texttt{rank}(\id(u),ID_{(c,u)})>|\texttt{bot}(I_{k,3}(u))|$ during the second round in phase $k+1$. Moreover, we can also partition $V_{k+1,3}(I_{k+1,3}(v))$ into sets $V^{unchanged}_{k+1,3}$ and $V^{halved}_{k+1,3}$.
	
	Similar to $B_{(c,u)}$, we define $T_{(c,u)}$ while committee member $c$ executing Figure~\ref{fig:alg-rename-committee-action} as follows:
	$$T_{(c,u)}=\{\id(w)~|~\langle\id(w),I_w,d_w,p_w\rangle\in M_c\text{ and }I_w\subseteq \texttt{top}(I_u)\}.$$
	
	One the one hand, by the definition of $T_{(c,u)}$, we have $V^{unchanged}_{k+1,3} \subseteq \{w\mid\id(w)\in T_{(c,u)}\}$.
	
	On the other hand, for any node $w\in V^{halved}_{k+1,3}$, it must in set $U$. Since $u$ has the smallest identifier in $U$, we have $\{\id(w)\mid w\in V^{halved}_{k+1,3}\} \subseteq \{i\mid i\in ID_{(c,u)}, i\geq\id(u)\}$.
	
	At this point, we have proved $|V_{k+1,3}(I_{k+1,3}(v))| \leq |T_{(c,u)}| + |\{i\mid i\in ID_{(c,u)}, i\geq\id(u)\}|$.
	
	By definition, we know that
	$$|T_{(c,u)}|=|\{\id(w)~|~\langle\id(w),I_w,d_w,p_w\rangle\in M_c\text{ and }I_w\subset I_u\}|-|B_{(c,u)}|,$$
	and that
	\begin{align*}
	& |\{i\mid i\in ID_{(c,u)}, i\geq\id(u)\}| \\
	= & |\{\id(w)~|~\langle\id(w),I_w,d_w,p_w\rangle\in M_c\text{ and }I_w=I_u\}|-\texttt{rank}(\id(u),ID_{(c,u)})+1.
	\end{align*}
	Therefore
	\begin{align*}
	& |V_{k+1,3}(I_{k+1,3}(v))| \\
	\leq & |T_{(c,u)}| + |\{i\mid i\in ID_{(c,u)}, i\geq\id(u)\}| \\
	\leq & |\{\id(w)~|~\langle\id(w),I_w,d_w,p_w\rangle\in M_c\text{ and }I_w\subseteq I_u\}|-(|B_{(c,u)}|+\texttt{rank}(\id(u),ID_{(c,u)}))+1 \\
	\leq & |\{\id(w)~|~\langle\id(w),I_w,d_w,p_w\rangle\in M_c\text{ and }I_w\subseteq I_u\}|-|\texttt{bot}(I_{k,3}(u))| \\
	\leq & |V_{k,3}(I_{k,3}(u))|-|\texttt{bot}(I_{k,3}(u))|.
	\end{align*}
	
	By the induction hypothesis, we know $|V_{k,3}(I_{k,3}(u))|\leq|I_{k,3}(u)|$. Hence, $|V_{k+1,3}(I_{k+1,3}(v))|\leq|I_{k,3}(u)|-|\texttt{bot}(I_{k,3}(u))|=|\texttt{top}(I_{k,3}(v))|=|I_{k+1,3}(v)|$. This competes the inductive step for the third round of phase $k+1$ for Scenario 1.2.
\end{itemize}

\smallskip\textsc{Scenario 2: $k'<k$.} In this scenario, $v$ does not halve its interval during phase $k+1$. That is, we have $I_{k+1,3}(v)=I_{k,3}(v)$.
For a node $u\neq v$, either $I_{k+1,3}(u)\cap I_{k+1,3}(v)=\emptyset$ or $I_{k+1,3}(u)\subseteq I_{k+1,3}(v)$. We further divide these two situations into four cases: (1) $I_{k+1,3}(u)\cap I_{k+1,3}(v)=\emptyset$; (2) $I_{k+1,3}(u)\subset I_{k+1,3}(v)$; (3) $I_{k+1,3}(u)=I_{k+1,3}(v)$ and $I_{k,3}(u)\neq\hat{I}_{k+1,3}(u)$; (4) $I_{k+1,3}(u)=I_{k+1,3}(v)$ and $I_{k,3}(u)=\hat{I}_{k+1,3}(u)$.

If $u$ is in case (1), clearly $u\notin V_{k+1,3}(I_{k+1,3}(v))$.

If $u$ is in case (2), since $I_{k+1,3}(v)=I_{k,3}(v)$, we have $u\in V_{k+1,3}(I_{k+1,3}(v))$ and $u\in V_{k,3}(I_{k,3}(v))$.

If $u$ is in case (3), then $u$ does not change its interval during phase $k+1$, hence again we have $u\in V_{k+1,3}(I_{k+1,3}(v))$ and $u\in V_{k,3}(I_{k,3}(v))$.

At this point, we know that if there does not exist a node $u$ belong to case (4), then $V_{k+1,3}(I_{k+1,3}(v)) \subseteq V_{k,3}(I_{k,3}(v))$. By the induction hypothesis, we conclude $|V_{k+1,3}(I_{k+1,3}(v))| \leq |V_{k,3}(I_{k,3}(v) | \leq |I_{k,3}(v)| = |I_{k+1,3}(v)|$, as required.

On the other hand, if there exists a node $u\neq v$ that belong to case (4), then we can apply the analysis of scenario one to node $u$, implying $|V_{k+1,3}(I_{k+1,3}(u))| \leq |I_{k+1,3}(u)|$. As a result, $|V_{k+1,3}(I_{k+1,3}(v))|=|V_{k+1,3}(I_{k+1,3}(u))|\leq|I_{k+1,3}(u)|=|I_{k+1,3}(v)|$, as required.

This competes the inductive step for the third round of phase $k+1$ for Scenario 2.
\end{proof}

\LemCrashRebuildCommittee*
\begin{proof}
By the end of phase $k$, if there does not exist any active node $v$ that has $elected$ equal to $true$, then all committee members have crashed. As a result, in phase $k+1$, Line~\ref{line:committee-decesion} of Figure~\ref{fig:alg-rename-main} will not be executed, and no node will receive any response from the committee during round three. Therefore, according to Line~\ref{line:no-message-received-from-committee} of Figure~\ref{fig:alg-rename-node-action}, every remaining active node will increase its $p_v$ by one. Thus, we have $\tilde{p}_{k+1,3}\geq\tilde{p}_{k,3}+1$.
\end{proof}

\LemBoundedDiffK*
\begin{proof}
By algorithm description, it is easy to verify that, for any node $v$, it only updates $p_v$ in Line~\ref{line:committee-change-pv} of Figure~\ref{fig:alg-rename-main}, and in Line~\ref{line:node-change-pv-1} or Line~\ref{line:node-change-pv-2} of Figure~\ref{fig:alg-rename-node-action}. Moreover, the first case is executed if and only if $elected_v$ equals $true$. We now prove this lemma by induction on $k$.

For the base case $k=0$, trivially $\hat{p}_{0,3}=\tilde{p}_{0,3}=0$, as required.

Suppose the claim holds for the $k$-th phase ($k\geq 0$), now consider phase $k+1$. Let $V'$ denote the set of active nodes at the end of phase $k+1$. Let $C$ denote the set of active committee members at the start of phase $k+1$. In the $(k+1)$-st phase, the nodes in $C$ either: (1) all have crashed before the start of round three; (2) at least one survives to the start of round three, but all have crashed before the end of phase $k+1$; or (3) at least one survives to the end of phase $k+1$. We consider these three scenarios separately:

\smallskip\textsc{Scenario 1.}
Define $\hat{p}'_{k,3}=\max_{v\in V'}p_{k,3}(v)$, $\tilde{p}'_{k,3}=\min_{v\in V'}p_{k,3}(v)$. By algorithm description, in round three of phase $k+1$, every remaining node $v$ executes Line~\ref{line:no-message-received-from-committee} of Figure~\ref{fig:alg-rename-node-action} and increases $p_v$ by one. Thus, we have $\hat{p}_{k+1,3}=\hat{p}'_{k,3}+1$ and $\tilde{p}_{k+1,3}=\tilde{p}'_{k,3}+1$. By the induction hypothesis, we have $\hat{p}'_{k,3}\leq\hat{p}_{k,3}\leq\tilde{p}_{k,3}+1\leq\tilde{p}'_{k,3}+1$. As a result, $\hat{p}_{k+1,3} = \hat{p}'_{k,3}+1 \leq \tilde{p}'_{k,3}+2 = \tilde{p}_{k+1,3}+1$, as required.

\smallskip\textsc{Scenario 2.}
According to the algorithm (specifically, Line~\ref{line:committee-change-pv} in Figure~\ref{fig:alg-rename-main}), by the end of round two of phase $k+1$, for each active committee member $c\in C$, it will update its $p$ value to the maximum among the messages it has received during this round.
As a result, we have $\max_{c\in C} p_{k+1,2}(c) \leq  \hat{p}_{k,3} \leq \tilde{p}_{k,3}+1$, where the last inequality is due to the induction hypothesis.
Now, we consider the following sub-scenarios:
\begin{itemize}
	\item Scenario 2.1: for every node $v\in V'$, it holds $p_{k+1,2}(v) = \tilde{p}_{k,3}$. In this case, for every node $v\in V'$, if it receives some response from the committee, then it updates its $p_v$ according to Line~\ref{line:node-change-pv-2} of Figure~\ref{fig:alg-rename-node-action}. Recall that we have shown $\max_{c\in C} p_{k+1,2}(c) \leq  \hat{p}_{k,3} \leq \tilde{p}_{k,3}+1$ above, thus $p_{k+1,3}(v)=\max\{p_{k,3}(v),\max_{c\in C} p_{k+1,2}(c)\}\leq\tilde{p}_{k,3}+1$. Otherwise, if $v$ does not receive any response from the committee, then it increases $p_v$ by one as in Line~\ref{line:node-change-pv-1} of Figure~\ref{fig:alg-rename-node-action}. Recall that we assume $p_{k+1,2}(v)=\tilde{p}_{k,3}$, thus $p_{k+1,3}(v)= \tilde{p}_{k,3}+1$. At this point, we have proved $\hat{p}_{k+1,3}\leq \tilde{p}_{k,3}+1$. On the other hand, trivially $\tilde{p}_{k+1,3}\geq\tilde{p}_{k,3}$. Therefore, we conclude $\hat{p}_{k+1,3}\leq \tilde{p}_{k+1,3}+1$ holds.
	
	\item Scenario 2.2: for every $v\in V'$, it holds $p_{k+1,2}(v)=\tilde{p}_{k,3}+1$. In this case, for every $c\in C$, we have $p_{k+1,2}(c)=\tilde{p}_{k,3}+1$. For any node $v\in V'$, if it receives some response from the committee, then it updates its $p_v$ according to Line~\ref{line:node-change-pv-2} of Figure~\ref{fig:alg-rename-node-action}, which leads to $p_{k+1,3}(v)=\max\{p_{k,3}(v),\max_{c\in C} p_{k+1,2}(c)\}=\tilde{p}_{k,3}+1$. Otherwise, if $v$ does not receive any response from the committee, then it increases $p_v$ by one as in Line~\ref{line:node-change-pv-1} of Figure~\ref{fig:alg-rename-node-action}, which leads to $p_{k+1,3}(v)=p_{k,3}(v)+1=\tilde{p}_{k,3}+2$. At this point, we have proved $\hat{p}_{k+1,3}\leq \tilde{p}_{k,3}+2$. On the other hand, since $p_{k+1,2}(v)=\tilde{p}_{k,3}+1$ for every node $v\in V'$, we know $\tilde{p}_{k+1,3}\geq\tilde{p}_{k,3}+1$. Therefore, we conclude $\hat{p}_{k+1,3}\leq \tilde{p}_{k+1,3}+1$ holds.
	
	\item Scenario 2.3: there exist two nodes $u,v\in V'$ such that $p_{k+1,2}(u)=\tilde{p}_{k,3}$ and $p_{k+1,2}(v)=\tilde{p}_{k,3}+1$. In this case, for every $c\in C$, again we have $p_{k+1,2}(c)=\tilde{p}_{k,3}+1$. Now, for a node $u\in V'$ with $p_{k+1,2}(u)=\tilde{p}_{k,3}$, regardless of whether it receives any response from the committee, it gets $p_{k+1,3}(u)=\tilde{p}_{k,3}+1$. On the other hand, for a node $v\in V'$ with $p_{k+1,2}(v)=\tilde{p}_{k,3}+1$, if it does not receive any response from the committee, then $p_{k+1,3}(v)=\tilde{p}_{k,3}+2$. Otherwise, if it receives some response from the committee, then $p_{k+1,3}(v)=\tilde{p}_{k,3}+1$. At this point, we have proved $\hat{p}_{k+1,3}\leq \tilde{p}_{k,3}+2$. Note that we assume there exists node $v\in V'$ with $p_{k+1,2}(v)=\tilde{p}_{k,3}+1$, thus $\tilde{p}_{k+1,3}\geq\tilde{p}_{k,3}+1$. Therefore, we conclude $\hat{p}_{k+1,3}\leq \tilde{p}_{k+1,3}+1$ holds.
\end{itemize}

\smallskip\textsc{Scenario 3.} In this case, every node $v\in V'$ receives some response from the committee in round three. By algorithm description, it is easy to verify that $\hat{p}_{k+1,3}\leq\hat{p}_{k,3}$. Thus, by the induction hypothesis, we conclude $\hat{p}_{k+1,3}\leq\hat{p}_{k,3}\leq\tilde{p}_{k,3}+1\leq\tilde{p}_{k+1,3}+1$ holds, as requited.
\end{proof}

\LemBoundedCimmitteeUpper*
\begin{proof}
Let $V$ be the set of all $n$ nodes. In this proof, for each node $v\in V$, let $p_{k,3}(v)$ denote the value of $p_v$ by the end of phase $k$ \highlightblue{even if $v$ has crashed by then}.

By our algorithm, after each phase, the value of $p_v$ either increases by one or remains unchanged. Thus, among the first $k$ phases, there are $p_{k,3}(v)$ phases in which the value of $p_v$ is increased. Moreover, by our algorithm, each time the value of $p_v$ is increased in some phase $k'\in[k]$, node $v$ will try to elect itself as a committee member with probability $(\highlightblue{256}\cdot2^{p_{k',3}(v)}\cdot\log{n})/n$ if it is not one already.

For each node $v\in V$, by the end of phase $k$, let $E_v$ be an indicator random variable that equals $1$ iff $elect_v$ \highlightblue{ever} is $true$. %\highlightblue{(This definition is valid even if $v$ has crashed by the end of phase $k$.)}
Now, an important observation is, for any fixed $(P_v)_{v\in V}\in[0,\hat{p}_{k,3}]^V$, conditioned on $p_{k,3}(v)=P_v$ for all $v\in V$, the set $\{E_v\mid v\in V\}$ is a set of mutually independent random variables.

Recall the discussion in the second paragraph of the proof, we have
$$\Pr[E_v=1] \leq 2\cdot(256\cdot2^{P_v}\cdot\log{n})/n \leq (2^{\hat{p}_{k,3}+9}\cdot\log{n})/n.$$
Let $E_V=\sum_{v\in V} E_v$, we have
$$\mathbb{E}[E_V]=\sum_{v\in V} \mathbb{E}[E_v]= \sum_{v\in V} \Pr[E_v=1]\leq 2^{\hat{p}_{k,3}+9}\cdot\log{n}.$$
Apply a Chernoff bound~\cite{motwani95}, we have $$\Pr[E_V\geq 3\mathbb{E}[E_V]]\leq e^{-\mathbb{E}[E_v]}\leq n^{-2^{\hat{p}_{k,3}+9}}\leq n^{-3}.$$

Note that $E_V$ is upper bounded by $n$ deterministically. Thus, conditioned on $p_{k,3}(v)=P_v$ for all $v\in V$, the number of nodes that \highlightblue{ever} have $elected_v$ equal to $true$ by the end of phase $k$ is at most $O(\min\{2^{\hat{p}_{k,3}}\cdot\log{n},n\})$, with probability at least $1-n^{-3}$.

Lastly, by the law of total probability, we conclude that the number of nodes that \highlightblue{ever} have $elected_v$ equal to $true$ by the end of phase $k$ is at most $O(\min\{2^{\hat{p}_{k,3}}\cdot\log{n},n\})$, with probability at least $1-n^{-3}$.
\end{proof}

\LemBoundedCimmitteeLower*
\begin{proof}
By algorithm description, it is easy to verify that the maximum $p_v$ value can increase by at most one after each phase. Thus, if $\hat{p}_{k,3}\geq3$, then among the first $k\geq3$ phases, there must exist at least three phases in which the maximum $p_v$ value among the remaining nodes increases by one. Assume phases $k_0$, $k_1$, and $k_2$ are the last three such phases. Specifically, assume that
\begin{align*}
& \hat{p}_{k_0-1,3}=\hat{p}_{k,3}-3,\quad\hat{p}_{k_0,3}=\hat{p}_{k,3}-2; \\
& \hat{p}_{k_1-1,3}=\hat{p}_{k,3}-2,\quad\hat{p}_{k_1,3}=\hat{p}_{k,3}-1; \\
& \hat{p}_{k_2-1,3}=\hat{p}_{k,3}-1,\quad\hat{p}_{k_2,3}=\hat{p}_{k,3}.
\end{align*}

Since $\hat{p}_{k_1,3}=\hat{p}_{k,3}-1$, by Lemma~\ref{lem:bounded-difference-k}, we know $\tilde{p}_{k_1,3}\geq\hat{p}_{k,3}-2$. Recall that $\hat{p}_{k_0-1,3}=\hat{p}_{k,3}-3$, hence from the beginning of phase $k_0$ to the end of phase $k_1$, every remaining node's $p_v$ value has increased by one.
Notice that by algorithm description, for any node that is not a committee member, whenever it increases its $p_v$ value, it will try to join the committee with probability $(256\cdot2^{p_v}\cdot\log{n})/n$.

Hence, if by the end of phase $k_1$, there are less than $n/2$ active nodes, then the adversary has already crashed at least $n/2\in\Omega(\min\{2^{\hat{p}_{k,3}}\cdot\log{n},n\})$ nodes, and we are done.

Otherwise, during phase $k_0$ to phase $k_1$, each of the at least $n/2$ active nodes will try to join the committee with probability at least $(256\cdot2^{\hat{p}_{k,3}-3}\cdot\log{n})/n$.
Let $C_1$ be the nodes among these active nodes that indeed join the committee. Then, we have
$$\mathbb{E}[|C_1|]\geq(n/2)\cdot (256\cdot2^{\hat{p}_{k,3}-3}\cdot\log{n})/n=2^{\hat{p}_{k,3}+4}\cdot\log{n}.$$
Since each node decides whether it joins the committee independently, apply a Chernoff bound, we have
$$\Pr[|C_1|\leq (1/2)\mathbb{E}[|C_1|]]\leq e^{-\mathbb{E}[|C_1|]/8}\leq n^{-2^{\hat{p}_{k,3}+1}}\leq n^{-3}.$$
Hence, by the end of phase $k_1$, there exist at least $|C_1|=\Omega(\min\{2^{\hat{p}_{k,3}}\cdot\log{n},n\})$ active committee members, with probability at least $1-n^{-3}$. Now consider phase $k_2$. Since the maximum $p_v$ value among the remaining nodes increases by one in this phase, we know Line~\ref{line:node-change-pv-1} of Figure~\ref{fig:alg-rename-node-action} has been executed by some node. This implies by the end of phase $k_2$, all committee members in $C_1$ must have crashed. Recall that $|C_1|=\Omega(\min\{2^{\hat{p}_{k,3}}\cdot\log{n},n\})$ with probability at least $1-n^{-3}$, the lemma is proved.
\end{proof}

\clearpage

\section{Pseudocode of the Byzantine-resilient Algorithm}\label{secapp:pseudocode-byz}

\begin{figure}[!h]
\hrule
\vspace{.5ex}
\textbf{The Byzantine-resilient renaming algorithm executed at each node $v$:}
\vspace{.3ex}
\hrule
\begin{small}
\begin{algorithmic}[1]
\State $committee_v\gets\emptyset$, $elected_v\gets false$, $\highlight{p_0\gets 8\log{n}/((1-3\epsilon_0)\epsilon_0^2n)}$, $\highlight{c_g}\gets (1-3\epsilon_0/2)(2/3+\epsilon_0)p_0n$.

\LineComment{Committee election and announcement.}
\State Construct $N$ binary random variables $(r_i)_{i\in [N]}\in \{0,1\}^{N}$ using shared randomness,
\Statex where each variable becomes $1$ \highlightblue{with probability $p_0$}.
\If{($r_{\id(v)}==1$)}
	\State $elected_v\gets true$.
	\State Send message $\langle ELECT, \id(v) \rangle$ to all nodes indicating $v$ is a committee member.
\EndIf
% \State $C_v\gets \{u\in V: p_{\id(u)}=1\} \cap \{u\in V: \text{$v$ receives a message $\langle ELECT, \id(u) \rangle$ from $u$}\}$ \Comment{Committee}
\For{(every message $\langle ELECT, \id(u) \rangle$ received $p_{\id(u)}=1$)}
	$committee_v\gets committee_v\cup\{\id(u)\}$. 
\EndFor

\LineComment{Original identity announcement.}
\State Send $\langle ID, \id(v) \rangle$ to all nodes in $committee_v$.
\If{($elected_v==1$)}
	\State Initialize $L_v$ to a vector of $N$ zeros.
	\For{(every received message $\langle ID, \id(u) \rangle$)}
		$L_v[\id(u)]\gets 1$.
	\EndFor
\EndIf

\LineComment{Committee reach consensus on $L_v$.}
\If{($elected_v==true$)}
	\State $J_v \gets \{\langle1,N\rangle\}$, \highlightblue{$\hat{J}_v\gets \emptyset$}, $dirty_v \gets \emptyset$. \Comment {Each element in $J_v$ is a tuple indicating an interval in $[N]$.}
	\While {($|J_v|>0$)}\label{line:byz-alg-while-loop}
		\State $j_v\gets J_v.\texttt{pop}()$.
		\If {($|j_v| > 1$)}
		\State Let $k=|j_v|$ and $\hash:\{0,1\}^k\rightarrow\{0,1\}^{\highlight{8\log{N}}}$ be a hash function constructed by shared randomness.
			%\State $s_v\gets\hash(L_v[j_v.\texttt{begin}(),\cdots,j_v.\texttt{end()}])$.
			%\State $cnt_v\gets\text{Number of 1s in }L_v[j_v.\texttt{begin}(),\cdots,j_v.\texttt{end()}]$.
			\State $s_v\gets\hash(L_v[j_v])$, $cnt_v\gets\text{number of 1s in }L_v[j_v]$.
			\State $\langle same_v,\highlightblue{\langle s'_v,cnt'_v\rangle}\rangle\gets$ run $\validator(\langle s_v,cnt_v\rangle)$ with nodes in $committee_v$.
			\State $same'_v\gets$ run $\consensus(same_v)$ with nodes in $committee_v$.
			\If{($same'_v==1$ \textbf{and} $\langle s_v,cnt_v\rangle\neq\highlightblue{\langle s'_v,cnt'_v\rangle}$)} $diff_v \gets 1$ \algorithmicelse\xspace $diff_v \gets 0$. \label{line:byz-set-diff}\EndIf
			\State Send $diff_v$ to all nodes in $committee_v$.
			\State $diff'_v \gets diff_v$. \label{line:byz-set-diffvprime-1}
			\If{(more than \highlight{$c_g/2$} nodes in $committee_v$ report $diff_u==1$)} $diff'_v \gets 1$. \label{line:byz-set-diffvprime-2} \EndIf
			\State $\highlightblue{diff''_v}\gets$ run $\consensus(diff'_v)$ with nodes in $committee_v$.
			\If{($same'_v==1$ \textbf{and} $\highlightblue{diff''_v}==0$)}
				\State \highlightblue{$\hat{J}_v\gets \hat{J}_v\cup \{j_v\}$}. 
				\If{($\langle s_v,cnt_v\rangle \neq \highlightblue{\langle s'_v,cnt'_v\rangle}$)\label{line:cond-not-dirty}}
					%\State Fill $L_v[j_v.\texttt{begin}(),\cdots,j_v.\texttt{end()}]$ with $\highlightblue{cnt'_v}$ $1$s arbitrarily. \label{line:byz-fill-cnt-1}
					\State Fill $L_v[j_v]$ with $\highlightblue{cnt'_v}$ $1$s arbitrarily. \label{line:byz-fill-cnt-1}
					\State $dirty_v \gets dirty_v \cup \{j_v\}$.
				\EndIf
			\Else
				%\State $J_v.\texttt{push}(j_v.\texttt{SecondHalf()})$, $J_v.push(j_v.\texttt{FirstHalf()})$.
				\State $J_v.\texttt{push}(\texttt{top}(j_v))$, $J_v.\texttt{push}(\texttt{bot}(j_v))$.
			\EndIf
		\Else
			\State \highlightblue{$\hat{J}_v\gets \hat{J}_v \cup \{j_v\}$}.
			\State $L_v[i]\gets$ run $\consensus(L_v[j_v])$ with nodes in $committee_v$.
		\EndIf
	\EndWhile
\EndIf

\LineComment {Obtain new identity.}
\If {($elected_v==true$)}
	\For {(every message $\langle ELECT, \id(u)\rangle$ received)}
		\If {($\id(u) \notin dirty_v$)}
			\State Send $\langle NEW,\nid(u)\rangle$ to node $u$ where $\nid(u)$ denotes the rank of $\id(u)$ in $L_v$.
		\Else
			\State Send $\langle NEW, \texttt{null} \rangle$ to node $u$.
		\EndIf
	\EndFor
\EndIf
\State \highlight{Wait for $c_g$ messages of type $NEW$ from $committee_v$.}
\State \highlight{Set $\id(v)$ be the majority value among all type $NEW$ messages received from $committee_v$.}
\end{algorithmic}
\end{small}
\hrule
\vspace{1ex}\caption{The Byzantine-resilient renaming algorithm executed at each node.}\label{fig:alg-rename-byz}\vspace{-4ex}
\end{figure}

\begin{figure}[!h]
\hrule
\vspace{.5ex}
\textbf{Procedure $\validator(in_v)$ executed at node $v$ with nodes in $members_v$:}
\vspace{.3ex}
\hrule
\begin{small}
\begin{algorithmic}[1]
\State Send $\langle INIT, in_v \rangle$ to all nodes in $members_v$.
\If{(the  most frequent value $in_u$ appears at least $\highlight{c_g}$ times in received $INIT$ messages)}
	\State Send $\langle ECHO, in_u \rangle$ to all nodes in $members_v$.
\EndIf
\State $out_v \gets in_v$, $same_v \gets 0$.
\State $in_1 \gets $ the most frequent value in received $ECHO$ messages. \State $in_2 \gets $ the second most frequent value in received $ECHO$ messages.
\If{($in_1$ appears more than $\highlight{c_g/2}$ times)} $out_v \gets in_1$. \EndIf
\If{((frequency difference of $in_1$ and $in_2$ is more than \highlight{$c_g/2$}) \textbf{and} ($in_1$ appears at least $\highlight{c_g}$ times))}
	\State $same_v \gets 1$.
\EndIf
\State \textbf{return} $\langle same_v,out_v\rangle$.
\end{algorithmic}
\end{small}
\hrule
\vspace{1ex}\caption{Procedure \validator\xspace executed at node $v$.}\label{fig:alg-validator}\vspace{-4ex}
\end{figure}

\section{Proofs for the Analysis of the Byzantine-resilient Algorithm}\label{secapp:analysis-byz}

\LemValidator*
\begin{proof}
Compared with the original algorithm \cite{lenzen22}, we allow the input to be a binary string instead of a single bit. To account for the fact that different nodes may have different views on the set of nodes that participate in the process (due to Byzantine nodes), we also make minor adjustments accordingly. See \Cref{fig:alg-validator} for the pseudocode. The following proof is self-contained.

We begin with part (1) of validity. Assume a correct node $v$ outputs $out_v$. Either $out_v=in_v$ or the value of $out_v$ appears more than $c_g/2$ times in $ECHO$ messages. In the former case we are done. In the latter case, since $c_g/2>|B|$, we know at least one correct node $w$ has sent $ECHO$ message for the value of $out_v$. This further implies $w$ has received at least $c_g$ $INIT$ messages containing the value of $out_v$. Again, since $c_g>|B|$, at least one such $INIT$ message originates from a correct node $u$, implying $out_v=in_u$.

Next, consider part (2) of validity. Assume $in_v=in$ for all $v\in G$. All correct nodes send same $INIT$ messages containing $in$, implying each correct node receives $INIT$ messages containing $in$ for at least $c_g$ times. Since $INIT$ messages containing values other $in$ can only come from Byzantine nodes and since there are less than $c_g/2$ Byzantine nodes, we know each correct node will only send $ECHO$ messages containing $in$. That is, $ECHO$ messages containing $in$ will be received by every correct nodes for at least $c_g$ times. Moreover, for any correct node, $ECHO$ messages containing values other than $in$ will appear less than $c_g/2$ times due to Byzantine nodes. Therefore, every correct node $v$ sets $out_v=in$ and $same_v=1$.

We continue with weak agreement.
Assume $same_v=1$ for some $v\in G$.
Let the most and the second most frequent value within $ECHO$ messages received by $v$  be $val_1$ and $val_2$ respectively.
Since $v$ sets $same_v$ to $1$, the frequency difference between $val_1$ and $val_2$ is more than $c_g/2$, and $v$ has received $val_1$ at least $c_g$ times.
Recall that there are less than $c_g/2$ Byzantine nodes, hence $val_1$ is the most frequent value in $ECHO$ messages for all correct nodes, and every correct nodes must have received more than $c_g/2$ $ECHO$ messages containing value $val_1$.
As a result, every correct node $u$ sets $out_u=val_1$, as required.

Lastly, note that each correct node sends at most $O(|G|+|B|)=O(\hat{c}_g)$ messages each of $O(\log{N})$ bits, and that there are $|G|=O(\hat{c}_g)$ correct nodes, the message and bit complexity follow.
\end{proof}

\LemConsensus*
\begin{proof}
Many algorithms could be used to achieve desired properties, here we use the \texttt{PolyByz} algorithm introduced in Section 6.3 of Lynch's classical textbook~\cite{lynch1996distributed} with a few minor adjustments. In particular, we use the lower bound $c_g$ on the number of correct nodes to replace the message number threshold $n-t$ (or equally, $2t+1$); we use $c_g/2$ to replace the message number threshold $t$; and we use the upper bound on the number of correct nodes $\hat{c}_g$ to bound the communication cost. With these adjustments, the analysis of \consensus\xspace is identical to that \texttt{PolyByz}, interested reader can refer to \cite{lynch1996distributed} for more details.
\end{proof}

\LemCommitteeAnnouncement*
\begin{proof}
According to algorithm description, once a correct node $v$ finds that $r_{\id(v)}=1$, it elects itself as a committee member and broadcasts its identifier $\id(v)$. Hence, every correct nodes $u$ will receives $v$'s identifier $\id(v)$, find $r_{\id(v)}=1$ and add $v$ to $C_u$. Thus, for any correct node $v$, it holds that $G\subseteq C_v$.

Next, we consider the bound on $|G|$. Recall that each correct node $v$ elects itself as a committee member iff $r_{\id(v)}=1$. Hence, $|G|=\sum_{v\text{ is a correct node}}r_{\id(v)}$. As a result, $\mathbb{E}[|G|]\geq(2/3+\epsilon_0)p_0n$ and $\mathbb{E}[|G|]\leq p_0n$. Since random variables $(r_i)_{i\in [N]}$ are sampled independently, apply a Chernoff bound, we know
\begin{align*}
\Pr[|G|\leq(1-3\epsilon_0/2)\mathbb{E}[|G|]] &\leq \exp(-9\epsilon_0^2\mathbb{E}[|G|]/8)\leq n^{-6},\\
\Pr[|G|\geq(1+3)\mathbb{E}[|G|]] &\leq \exp(-9\mathbb{E}[|G|]/5)\leq n^{-6}.
\end{align*}
That is, with probability at least $1-2/n^6$, it holds that $c_g\leq|G|\leq\hat{c}_g$.

Lastly, we consider the bound on $|B|$. Recall that messages are authenticated, hence all nodes cannot fake their identifiers. Thus, $|B|\leq\sum_{v\text{ is a Byzantine node}} r_{\id(v)}$, implying $\mathbb{E}[|B|]\leq(1/3-\epsilon_0)p_0n$. Since random variables $(r_i)_{i\in [N]}$ are sampled independently after the adversary has determined which nodes are Byzantine, apply a Chernoff bound, we know
$$\Pr[|B|\geq(1+3\epsilon_0)\mathbb{E}[|B|]]\leq \exp(-3\epsilon_0^2\mathbb{E}[|B|]) \leq n^{-8}.$$

Apply a union bound, the lemma is immediate.
\end{proof}

\LemSameJv*
\begin{proof}
We prove the claim via induction on iterations. (That is, the \textbf{while} loop at Line~\ref{line:byz-alg-while-loop} in \Cref{fig:alg-rename-byz}.) For the induction basis, every correct committee member $v$ initializes $J_v$ to $\{\langle 1,n\rangle\}$ and $\hat{J}_v$ to the empty set.

Suppose the claim holds at the end of the $i$-th iteration and the \textbf{while} loop does not terminate at the end of iteration $i$, we now show the claim holds at the end of the $(i+1)$-st iteration.

By the induction hypothesis, at the end of iteration $i$, all correct committees have identical $J_v$. Thus, in iteration $i+1$, all correct committees work on identical interval $j_v$.

Suppose $j_v = [l,r]$. Depending on the size of $j_v$, there are two cases.

If $|j_v|=1$, every correct committee member $v$ runs a consensus algorithm (particularly, \consensus) on $L_v[j_v]$ and adds $j_v$ to $\hat{J}_v$ in iteration $i+1$. Thus, at the end of iteration $i+1$, all correct committee members have identical $J_v$, $\hat{J}_v$; and $J_v\cup \hat{J}_v$ is a partition of $[N]$.

If $|j_v|>1$, the modification on $J_v$ and $\hat{J}_v$ depends on the value of $same_v'$ and $diff_v'$, resulting from two executions of the consensus algorithm \consensus.
By \Cref{lem:consensus}, assuming \Cref{item:cond-byz-1} of \Cref{cond:byz}, all correct committees have identically valued $same_v'$ and $diff_v''$.
Hence, if $same_v'=1$ and $diff_v''=0$ for a correct committee $v$, then all correct committee members have $same_v'=1$, $diff_v''=0$ and add $j_v$ to $\hat{J}_v$.
Otherwise, all correct committee members add the first and the second half of $j_v$ to $J_v$.

As a result, when $|j_v|>1$, by the end of iteration $i+1$, again all correct committee members have identical $J_v$, $\hat{J}_v$ and $J_v\cup \hat{J}_v$ is a partition of $[N]$. This completes the proof for the inductive step.
\end{proof}

\LemSamejv*
\begin{proof}
By \Cref{lem:while-same-Jv}, assuming \Cref{item:cond-byz-1} of \Cref{cond:byz}, at the end of every iteration of the \textbf{while} loop, all correct committee members $v$ have the same $J_v$, $\hat{J}_v$, and $J_v\cup \hat{J}_v$ is a partition of $[N]$. Thus, in any iteration, all correct committee member works on the same $j_v$. Consider an arbitrary iteration $i$, assume all correct committee members have identically-valued $L_v[j_v]$. Depending on the size of $j_v$, there are two cases.

If $|j_v|=1$, all correct committee members run a consensus algorithm \consensus\xspace to reach agreement on $L_v[j_v]$ and add $j_v$ to $\hat{J}_v$ at the end of iteration $i$.

If $|j_v|>1$, all correct committee members have identically valued $\langle s_v, cnt_v\rangle$.
By \Cref{lem:validator}, assuming \Cref{item:cond-byz-1} of \Cref{cond:byz}, the output of \validator---i.e., $\langle same_v, \langle s_v', cnt_v'\rangle \rangle$---is identical for all correct committee members. Moreover, due to the validity property of \validator, $same_v=1$ and $\langle s_v', cnt_v'\rangle =\langle s_v, cnt_v\rangle $.
By \Cref{lem:consensus}, assuming \Cref{item:cond-byz-1} of \Cref{cond:byz}, after the consensus process of $same_v$, all correct committee members output $same_v'=1$.
That is, every correct committee member $v$ has $same_v'=1$ and $\langle s_v', cnt_v'\rangle =\langle s_v, cnt_v\rangle $, and sets $diff_v=0$.
Hence, assuming \Cref{item:cond-byz-1} of \Cref{cond:byz}, after committee members broadcast $diff_v$, no correct committee member can receive more than $c_g/2$ messages signaling $diff_v=1$.
As a result, all correct committee members set $diff_v'=0$.
At this point, By \Cref{lem:consensus}, assuming \Cref{item:cond-byz-1} of \Cref{cond:byz}, all correct committee members output $diff_v''=0$.
We conclude, at the end of iteration $i$, all correct committee members have $same_v'=1$ and $diff_v''=0$, and add $j_v$ to $\hat{J}_v$.
\end{proof}

\LemWhileLoopCount*
\begin{proof}
Imagine a binary tree $T$, in which each vertex is associated with an interval.
For each vertex $x\in T$, denote the interval associated with it as $I_x$.
We construct $T$ in the following manner.
The root of $T$ is labeled with interval $[1,N]$.
When committee members process the segment $L[I_x]$ which corresponds to interval $I_x$ associated with vertex $x\in T$, if $|I_x|=1$ or if all correct committee members agree on $L_v[I_x]$ (i.e., $same'_v=1$ and $diff''_v=0$ for all correct committee members in that iteration), then vertex $x$ is a leaf of $T$.
Otherwise, $x$ has two children: left child is labeled with $\texttt{bot}(I_x)$, right child is labeled with $\texttt{top}(I_x)$.

By \Cref{lem:while-same-jv}, the number of iterations of the \textbf{while} loop can be bounded by the number of vertices in $T$. To that end, note that the label associated with an vertex is halved whenever we move alone an edge against the root.
Thus, the depth $d$ of binary tree $T$ is less than $2\log{N}$.
For any $i\in[d]$, the labels of the vertices at depth $i$ must be a subset of a partition of $[n]$.
Recall there are $f$ Byzantine nodes. Since nodes cannot fake their identifiers, correct committee members will disagree on at most $f$ positions in their identity lists (which is a size $N$ bit vector) after original identity announcement.
Therefore, there are at most $f$ vertices at depth $i$ whose labels correspond to segments that are not agreed upon by all correct committee members.
This implies, for any $i\in [d]$, there are at most $2f$ vertices with depth $i+1$.
In conclusion, there are at most $d\cdot 2f=4f\cdot \log N$ vertices in binary tree $T$.
\end{proof}

\LemMajorityCommitteeConsensus*
\begin{proof}
We prove the lemma via induction on iterations. For the induction basis, every correct committee member $v$ initializes $\hat{J}_v$ to an empty set. Thus, the claim holds trivially.

Suppose the claim holds up to the end of iteration $i$, now consider iteration $i+1$.
Note that in iteration $i+1$, the algorithm will not work on any intervals that have already been processed, thus we focus on the interval being processed in iteration $i+1$. By \Cref{lem:while-same-jv}, in iteration $i+1$, all correct committee members work on the same interval. Consider an arbitrary correct committee member $v$, suppose $v$ (hence all correct committee members) works on $j_v=[l,r]$.
Depending on the size of $j_v$, there are two scenarios.

\smallskip\textsc{Scenario I}: $|j_v|=1$.
In this scenario, every correct committee member $v'$ runs \consensus\xspace on $L_{v'}[j_v]$ and adds $j_v$ to $\hat{J}_{v'}$.
By \Cref{lem:consensus}, assuming \Cref{item:cond-byz-1} of \Cref{cond:byz}, this execution of \consensus\xspace succeeds.
Hence, part (a) and part (b) of \Cref{item:majority-1} hold.
According to \Cref{fac:announce}, for the single identity $i\in j_v$, if there is a correct node whose original identity is $i$, then before running \consensus\xspace for $L_v[j_v]$, every correct committee member $v'$ has input $L_{v'}[i]=1$; moreover, after running \consensus\xspace for $L_v[j_v]$, every correct committee member outputs $L_{v'}[i]=1$, due to \Cref{lem:consensus}.
Hence, part (c) of \Cref{item:majority-1} holds.
At this point, it is ease to see that \Cref{item:majority-2} also holds.

\smallskip\textsc{Scenario II}: $|j_v|>1$.
In this scenario, the modification to $\hat{J}_v$ depends on the value of $same_v'$ and $diff_v''$, which are the results of two executions of procedure \consensus.
By \Cref{lem:consensus}, assuming \Cref{item:cond-byz-1} of \Cref{cond:byz}, these two executions succeed. Thus, every correct committee member $v'$ has identically valued $same_{v'}'$ and $diff_{v'}''$.
If $v$ finds $same_{v}' \neq 1$ or $diff_{v}'' \neq 0$, then it will not add $j_v$ to $\hat{J}_v$. Moreover, all correct committee member will do so as well. In this case, the claim holds trivially. So, in the following analysis, assume $v$ finds $same_v'=1$ and $diff_v''=0$.

Recall that correct committee member $v$ runs \consensus\xspace with the committee in its view with input $same_v$ and outputs $same_v'$.
Thus, due to the validity property of \consensus\xspace stated in \Cref{lem:consensus}, if $same_v'=1$, there must exist a correct committee member $u_1$ with input $same_{u_1}=1$.
This implies $u_1$ runs \validator\xspace with the committee in its view with input $\langle s_{u_1}, cnt_{u_1}\rangle$ and outputs $\langle same_{u_1}, \langle s_{u_1}', cnt'_{u_1}\rangle \rangle$.
By \Cref{lem:validator}, assuming \Cref{item:cond-byz-1} of \Cref{cond:byz}, this execution of \validator\xspace succeeds.
Moreover, since $same_{u_1}=1$, there must exist a correct committee member $u_2$ such that, for any correct committee member $v'$, it holds that $\langle s_{v'}', cnt_{v'}'\rangle=\langle s_{u_1}', cnt_{u_1}'\rangle=\langle s_{u_2}, cnt_{u_2}\rangle$.
Here, the first equation is due to the weak agreement property of \validator\xspace and the second equation is due to the validity property of \validator.
%At this point, we conclude that all correct committee members reach agreement on the number of $1$s in $L_v[j_v]$. (Particularly, this count is $cnt_{u_2}$.)

Recall that correct committee member $v$ runs \consensus\xspace with the committee in its view with input $diff_v'$ and outputs $diff_v''=0$.
By \Cref{lem:consensus}, assuming \Cref{item:cond-byz-1} of \Cref{cond:byz}, this execution of \consensus\xspace succeeds.
Thus, due to the validity property of \consensus, since $v$ has $diff_v'' = 0$, there must exist a correct committee member $u_3$ with $diff_{u_3}'=0$.
By our algorithm (particularly, Line~\ref{line:byz-set-diffvprime-1} and Line~\ref{line:byz-set-diffvprime-2} of \Cref{fig:alg-rename-byz}), if $diff_{u_3}'=0$, then $diff_{u_3}=0$ and at most $c_g/2$ committee members in the view of $u_3$ report $1$-valued $diff$ to $u_3$.
%Call this set of (not necessarily all correct) committee members as $W$.
By \Cref{item:cond-byz-1} of \Cref{cond:byz}, there are at least $c_g$ good committee members; moreover, by \Cref{lem:committee-announcement}, these at least $c_g$ good committee members are in the view of $u_3$.
Therefore, among the committee members in the view of $u_3$, at least $c_g/2$ of them report $0$-valued $diff$ to $u_3$ and they are correct nodes. Call this set of correct committee members as $G'$.
Note that by our algorithm (particularly, Line~\ref{line:byz-set-diff} of \Cref{fig:alg-rename-byz}), for each node $x\in G'$, value of $diff_x$ is $1$ iff $same_x'=1$ and $\langle s_x,cnt_x\rangle\neq\langle s_x', cnt_x'\rangle$. Moreover, recall that for each node $x\in G'$, it holds that $same'_x=1$.
Hence, for each node $x\in G'$, it holds that $\langle s_x,cnt_x\rangle=\langle s_x', cnt_x'\rangle$.

As a result, by our algorithm (particularly, Line~\ref{line:cond-not-dirty} of \Cref{fig:alg-rename-byz}), each node $x\in G'$ will not mark $j_v$ as dirty, proving part (a) of \Cref{item:majority-1} in the lemma statement. Recall the analysis in the second to last paragraph, for each node $x\in G'$ it holds that $\langle s_x',cnt_x'\rangle=\langle s_{u_2}, cnt_{u_2}\rangle$. Assuming \Cref{item:cond-byz-3} of \Cref{cond:byz}, this implies for each node $x\in G'$ it holds that $L_x[j_x]=L_{u_2}[j_{u_2}]$ at the beginning and the ending of iteration $i+1$, proving part (b) of \Cref{item:majority-1} in the lemma statement. Note that for each node $x\in G'$, segment $L_x[j_x]$ remains unchanged in iteration $i+1$. Recall \Cref{fac:announce}, part (c) of \Cref{item:majority-1} in the lemma statement is proved.

Lastly, consider \Cref{item:majority-2} in the lemma statement. By above analysis, all correct committee members reach agreement on the number of $1$s in $L_v[j_v]$. (Particularly, this count is $cnt_{u_2}$.) Moreover, this count equals the actual number of $1$s in $L_{u_2}[j_{u_2}]$ at the beginning of iteration $i+1$.
At this point, by \Cref{fac:announce}, \Cref{item:majority-2} in the lemma statement is proved.
\end{proof}

\LemByzIDUnique*
\begin{proof}
%By \Cref{lem:while-loop-iterations}, the \textbf{while} loop in the algorithm terminates in $O(f\cdot \log N)$ iterations.
By \Cref{lem:while-same-Jv}, at the end of the last \textbf{while} iteration, every correct committee member $v$ finds $J_v$ is an empty set and has identical $\hat{J}_v$, which is a partition of $[N]$.
Let $k=|\hat{J_v}|$. Sort the intervals in $\hat{J_v}$ by their left endpoints in the ascending order, denote the resulting intervals as $I_1,I_2, \cdots, I_k$.
Consider an arbitrary interval $I_j$ where $1\leq j\leq k$.
By \Cref{item:majority-2} of \Cref{lem:majority-committee-consensus}, all correct committee members reach the consensus on the number of $1$s in the segment corresponding to $I_j$.
Denote the number of 1s in $L_v[I_j]$, where $v$ is a correct committee member, by $cnt_j$.
Denote the number of nodes with identifiers in $I_j$ by $a_j$. By \Cref{item:majority-2} of \Cref{lem:majority-committee-consensus}, $cnt_j\leq a_j$.
By \Cref{item:majority-1} of \Cref{lem:majority-committee-consensus}, there exists a set $G'$ containing at least $c_g/2$ correct committee members such that: (1) every $v\in G'$ does not mark $I_j$ as dirty; (2) every $v\in G'$ agrees on $L_v[I_j]$; and (3) for any $i\in [I_j]$, if $i$ is the original identity of a correct node, then $L_v[i]=1$ for every $v\in G'$.

Now, consider an arbitrary $i_i\in [I_j]$ such that $i_u$ is the original identity of a correct node $u$. By above analysis, there exists a set $G'$ containing at least $c_g/2$ correct committee members such that every $v\in G'$ sends message $\langle NEW,\texttt{rank}(i_u,L_v)\rangle$ to $u$.
Here, $\texttt{rank}(i_u,L_v)$ denotes the number of $1$s in $L_v$ that occurs before position $i_u$.
Moreover, every correct committee member not in $G'$ will send $\langle NEW, \texttt{null}\rangle$ to $u$. Recall that $|G|\geq c_g$ and $|B|<c_g/2$. As a result, node $v$ will set $\texttt{rank}(i_u,L_v)$ as its new identity.

It is easy to see all correct nodes obtain distinct new identities and this renaming is order-preserving.
Moreover, since $\sum_{j\in [k]} cnt_j\leq \sum_{j\in [k] }a_j=n$, this renaming is strong.
\end{proof}

\section{Proof for the Lower Bound}\label{secapp:lower-bound}

\ThmLowerBound*

\begin{proof}
As mentioned earlier, our high-level strategy for deriving the above bound is to first define an anonymous version of the renaming problem in which nodes do not hold initial identities; next, we leverage the fact that if nodes exchange too few messages then two of them will have colliding new identities in $[n]$ with non-trivial probability, hence obtaining a lower bound for the anonymous renaming problem; finally, we reduce the anonymous renaming problem to its regular version, arriving at the desired result.

We now proceed to the proof, and begin with a formal definition for the anonymous renaming problem.

\begin{definition}[Anonymous renaming]
Consider a distributed system consisting of $n$ nodes. In the \emph{anonymous renaming} problem, each node $v$ needs to select a unique identity $\nid(v)\in[M]$, where $n\leq M$. That is, for any two nodes $u\neq v$, it holds that $\nid(u)\neq\nid(v)$.
\end{definition}

In this proof, we assume every node knows the node count $n$ and can independently generate private random bits. We also assume nodes can generate shared random bits and all messages exchanged are authenticated. Note that we focus on strong renaming, hence $M=n$ throughout the proof.

Next, we show that in the above setting, for any anonymous renaming problem succeeding with probability at least $5/8$, it must send more than $n/256$ messages in expectation.

To see this, suppose that $\mathcal{A}$ is a strong anonymous renaming algorithm that sends at most $n/256$ messages in expectation. By Markov's inequality~\cite{motwani95}, the probability that $\mathcal{A}$ sends more than $n/64$ messages is at most $1/4$. In the following, we focus on executions of $\mathcal{A}$ in which at most $n/64$ messages are sent.

Consider an arbitrary such execution, since at most $n/64$ messages are sent, by the pigeonhole principle, there are at least two nodes, denoted by $v_1$ and $v_2$, that send or receive no messages throughout the execution. In the following, we consider the executions in which $v_1$ and $v_2$ send or receive no messages. For any node $v$, let $\mathcal{A}(v)$ denote the final output of node $v$, which is $v$'s new identity in our setting.
Recall that nodes do not have initial identities. By an argument of symmetry, for any $i\in[n]$, we have
$$\Pr[\mathcal{A}(v_1)=i]=\Pr[\mathcal{A}(v_2)=i],$$
which means $v_1$ and $v_2$ share the same probability to select an identity. Note that it holds even if $v_1$ and $v_2$ have access to shared randomness and the messages exchanged between other nodes are authenticated.

Let $V$ denote the set of all nodes. For a node set $V'\subseteq V$, let $\mathcal{A}(V')$ denote the set $\{\mathcal{A}(v)\mid v\in V' \text{ and $v$ does not fail throughout execution}\}$ and let $K_{\mathcal{A}}(V')$ denote the event that nodes in $V'$ choose distinct identities. Additionally, we use $\hat{V}=V\setminus\{v_1,v_2\}$ to denote the node set without $v_1$ and $v_2$.

Suppose the adversary does not corrupt any node. In that case, there are three situations for $\mathcal{A}$ to fail: (1) at least two nodes in $\hat{V}$ choose the same identity; (2) situation (1) does not occur, but $v_1$ or $v_2$ chooses the same identity as some node in $\hat{V}$; (3) situation (1) and (2) do not occur, but $v_1$ and $v_2$ choose same identity. Therefore, in case the adversary does not corrupt any node, algorithm $\mathcal{A}$ will fail with probability
\begin{align*}
\Pr[\mathcal{A}\text{ fails}] = \Pr[\neg K_{\mathcal{A}}(V)] = &\Pr[\neg K_{\mathcal{A}}(\hat{V})]~+ \\
&\Pr[( \mathcal{A}(v_1) \in \mathcal{A}(\hat{V})) \vee (\mathcal{A}(v_2) \in \mathcal{A}(\hat{V})) \mid K_{\mathcal{A}}(\hat{V})]~+ \\
&\Pr[\mathcal{A}(v_1)=\mathcal{A}(v_2)\mid (\mathcal{A}(v_1),\mathcal{A}(v_2)\notin\mathcal{A}(\hat{V})) \wedge K_{\mathcal{A}}(\hat{V})].
\end{align*}

We are interested in the failure probability caused by $v_1$ and $v_2$, i.e., the failure probability in situations (2) and (3). To that end, let $p_i$ denote the probability that $v_1$ chooses $i$. That is, $p_i=\Pr[\mathcal{A}(v_1)=i]$. Then,
$$\Pr[\mathcal{A}(v_1) \in \mathcal{A}(\hat{V}) \vee \mathcal{A}(v_2) \in \mathcal{A}(\hat{V}) \mid K_{\mathcal{A}}(\hat{V})] \geq \Pr[ \mathcal{A}(v_1) \in \mathcal{A}(\hat{V}) \mid K_{\mathcal{A}}(\hat{V})] = \sum_{i\in\mathcal{A}(\hat{V})}p_i$$
and
\[\Pr[\mathcal{A}(v_1)=\mathcal{A}(v_2) \mid (\mathcal{A}(v_1),\mathcal{A}(v_2)\notin\mathcal{A}(\hat{V})) \wedge K_{\mathcal{A}}(\hat{V})] = \sum_{i\in [n]\setminus \mathcal{A}(\hat{V})} p_i^2.\]
As a result, we know
$$\Pr[\mathcal{A}\text{ fails}] \geq \sum_{i\in\mathcal{A}(\hat{V})}p_i + \sum_{i\in[n]\setminus\mathcal{A}(\hat{V})}p_i^2 = 1+\sum_{i\in[n]\setminus\mathcal{A}(\hat{V})}(p_i^2-p_i).$$

It is easy to verify that $\sum_{i\in[n]\setminus\mathcal{A}(\hat{V})}(p_i^2-p_i)$ attains minimum value when $p_i={1}/{|[n]\setminus\mathcal{A}(\hat{V})|}$ for every $i\in[n]\setminus\mathcal{A}(\hat{V})$. Moreover, as $\sum_{i\in [n]\setminus\mathcal{A}(\hat{V})}p_i$ gets bigger, the minimum value of $\sum_{i\in[n]\setminus\mathcal{A}(\hat{V})}(p_i^2-p_i)$ gets smaller. Also, recall that in situations (2) and (3), $|[n]\setminus \mathcal{A}(\hat{V})|=2$. Therefore, the minimum value that $\sum_{i\in[n]\setminus\mathcal{A}(\hat{V})}(p_i^2-p_i)$ could possibly attain is reached when $p_i=1/2$ for all $i\in [n]\setminus\mathcal{A}(\hat{V})$ and $p_i=0$ for all $i\in\mathcal{A}(\hat{V})$. This implies $\Pr[\mathcal{A}\text{ fails}]\geq 1/2$.

At this point, we can conclude, for any randomized anonymous renaming problem succeeding with probability at least $1-(1-1/4)\cdot(1/2)=5/8$, it must send more than $n/256$ messages in expectation.

The last step before we arrive at the theorem is to apply a reduction argument. Specifically, suppose $\mathcal{B}$ is a standard renaming algorithm that sends at most $n/256$ messages in expectation and succeeds with probability at least $3/4$. We can construct an algorithm $\mathcal{A}'$ that solves the anonymous renaming problem as follows: for each node $v$, it first chooses an identity $\id(v)$ from $[1,N]$ uniformly at random, then node $v$ runs algorithm $\mathcal{B}$ with $\id(v)$ as its input. Clearly, for any two nodes $v_1$ and $v_2$, we know $\Pr[\id(v_1)=\id(v_2)]=1/N$. Apply a union bound, the probability that any two nodes have colliding $\id$ is at most ${n\choose 2}\cdot(1/N)\leq n^2/(2N)\leq 1/10$. Therefore, the probability that $\mathcal{A}'$ fails is at most $(1-3/4)+1/10<3/8$. That is, $\mathcal{A}'$ is a randomized anonymous renaming algorithm succeeding with probability at least $5/8$ and sends at most $n/256$ messages in expectation, resulting in a contradiction with our above lower bound.

Finally, note that any lower bound on the number of messages immediately implies the same lower bound on the number of bits, hence the theorem is proved.
\end{proof}

\end{document}